\documentclass{article}
\usepackage[ruled,vlined]{algorithm2e}

\usepackage{arxiv}

\usepackage[utf8]{inputenc} 
\usepackage[T1]{fontenc}    
\usepackage{hyperref}       
\usepackage{url}            
\usepackage{booktabs}       
\usepackage{amsfonts}       
\usepackage{nicefrac}       
\usepackage{microtype}      
\usepackage{lipsum}
\usepackage{graphicx}
\graphicspath{ {./images/} }
\usepackage{amsmath} 
\usepackage{enumitem}

\usepackage{booktabs}
\usepackage{threeparttable}
\usepackage{graphicx}
\usepackage[table]{xcolor}
\usepackage{amsmath}
\usepackage{caption}

\usepackage[
backend=biber,
style=numeric,
doi=true,
url=true,
eprint=true,
isbn=false,
sorting=none
]{biblatex}

\title{Programmable Silicon Retina on Pixel Processor Array}

\author{
 Maciej Lewandowski \\
  Department of Electrical and Electronic Engineering \\
  University of Manchester, UK\\
  \texttt{maciej.lewandowski@postgrad.manchester.ac.uk} \\
   \And
 Prince Philip \\
   Department of Electronic Systems Engineering,\\ Indian Institute of Science, Bangalore, India\\
   \texttt{princephilip@iisc.ac.in} \\
  \And
   Alexandre Marcireau \\
  Department of Computer Science \\
  University of Manchester, UK\\
 \texttt{alexandre.marcireau@manchester.ac.uk} \\
  \And
  Chetan Singh Thakur \\
  Department of Electronic Systems Engineering \\
  Indian Institute of Science, Bangalore, India\\
 \texttt{csthakur@iisc.ac.in} \\
  \And
  André van Schaik \\
  Department of Computer Science \\
  University of Manchester, UK\\
 \texttt{andre.vanschaik@manchester.ac.uk} \\
  \And
 Piotr Dudek \\
  Department of Electrical and Electronic Engineering \\
  University of Manchester, UK\\
  \textbf{corresponding author} \\ 
  \textbf{p.dudek@manchester.ac.uk} \\
}

\begin{document}
\maketitle
\begin{abstract}

Standard dynamic vision sensors approximate retinal processing by detecting temporal contrast changes, offering high speed and high dynamic range. In this work, we explore whether incorporating additional biologically inspired processing stages - specifically spatial filtering and gain control - can offer advantages for certain downstream tasks such as saliency prediction. We present the first implementation of a multi-stage Silicon Retina model on the SCAMP-5 Pixel Processor Array, along with a GPU-based simulation framework. We evaluate the performance of our model on Video Intensity Reconstruction and Video Saliency Prediction. While the bio-inspired model is less effective at reconstructing absolute intensity frames, it achieves a 13\% reduction in saliency prediction loss in comparison to standard DVS event representation, while reducing the event rate by approximately 47\%. These experiments are obtained using a lightweight $\approx 100$k-parameter FireNet-style network, adapted from event-based reconstruction to saliency prediction. These results suggest that the silicon retina's "information distillation" mechanism can achieve a more efficient representation for downstream neural networks, particularly in bandwidth-constrained edge applications.

\end{abstract}


\section{Introduction}
Conventional frame-based CMOS image sensors have achieved remarkable advances in resolution and cost efficiency. However, they remain poorly suited for high-speed (>1k FPS) or low-power vision applications \cite{maqueda_event-based_2018, scheerlinck_fast_2020, magrini_drone_nodate, gallego_event-based_2022}. 
Streaming full-resolution frames at kilohertz rates imposes severe bandwidth and latency constraints, often requiring energy-intensive GPUs to process the vast amount of data. In resource-constrained environments, such as high-speed robotics, these limitations often require trade-offs, such as frame downsampling or algorithmic simplification, which introduces latency or degrades performance. Furthermore, for battery-constrained systems such as monitoring or video analytics, analysing static scenes is often inefficient, as the primary information of interest lies in dynamic changes and movement. 

In contrast, Dynamic Vision Sensors (DVS) offer a biologically inspired alternative \cite{posch_retinomorphic_2014, gallego_event-based_2022, philip_neuromorphic_2025}. Rather than capturing complete frames, these sensors encode visual input as sparse, asynchronous streams of brightness changes (polarity events) \cite{gallego_event-based_2022,chakravarthi_recent_2024}. When transmitted sparsely, these event streams inherently minimise bandwidth and latency, offering high temporal resolution. Consequently, event cameras have been successfully applied in domains including space imaging \cite{ralph_astrometric_2023}, satellite observation \cite{arja_noise_2024}, high-speed robotics \cite{gallego_event-based_2022, li_robust_2019, bardow_simultaneous_2016}, autonomous driving \cite{maqueda_event-based_2018}, and high-framerate video reconstruction \cite{rebecq_high_2019, scheerlinck_fast_2020}. Furthermore, the asynchronous nature of events aligns naturally with emerging neural architectures such as Spiking Neural Networks \cite{gruel_bio-inspired_2021, arfa_efficient_2025} or asynchronous event-based graph neural networks \cite{schaefer_aegnn_2022, yang_evgnn_2025}. Assuming no noise, it is possible to recover full frames from events using simple filtering or classical methods \cite{scheerlinck_continuous-time_2018, bardow_simultaneous_2016}. 

Most commercially available event cameras \cite{noauthor_metavision_nodate,noauthor_buy_nodate,prophesee_paper} implement a relatively simplified model of retinal processing. They primarily mimic thresholded photoreceptor derivatives, and skip key biological features such as spatial filtering, contrast gain control and retinal adaptation mechanisms \cite{carver_mead_silicon_nodate,wohrer_virtual_2009,field_information_2007}. To address the need for more sophisticated in-sensor processing, Pixel-Processor Arrays (PPAs) offer a fundamentally different paradigm. PPAs integrate programmable Processing Elements (PEs) directly into the imaging array, integrating sensing and processing at the pixel level \cite{dudek_sensor-level_2022, carey_100000_2013}. Similar to DVS, PPAs can reduce power consumption by transmitting sparse feature data instead of raw frames. PPAs also enable the execution of complex visual pipelines, such as feature tracking \cite{bose_descriptor--pixel_nodate}, gaze tracking \cite{bose_pixel_2022} or running a small neural network creating an embedding \cite{so_pixelrnn_2023} —all at kilohertz frame rates with sub-watt power budgets. 

In this work, we present an implementation of a biologically inspired silicon retina model on the SCAMP-5 PPA \cite{dudek_sensor-level_2022, carey_100000_2013}. Unlike standard event sensors, our implementation captures a richer approximation of retinal dynamics, incorporating stages that resemble the Outer Plexiform Layer (OPL), Contrast Gain Control and spiking Ganglion Cells \cite{wohrer_virtual_2009, philip_neuromorphic_2025}. We map these stages onto the SCAMP-5 hardware using analog operations and superpixel memory layout \cite{martel_pixel_interlacing_2015}.
Deploying such a complex model on SCAMP-5 poses significant challenges due to per-pixel memory limits and arithmetic noise. Given these constraints and the limited commercial availability of PPA hardware, we also provide a complementary digital simulation framework. This setup utilises a high-speed commercial image sensor and custom CUDA kernels to emulate the retinal model on a GPU. 


Having demonstrated that the proposed silicon retina can be implemented both on analog hardware (PPA) and in a CUDA-based simulation framework, we next investigate whether its additional biological complexity provides any advantage over a standard DVS representation in downstream vision tasks. We therefore compare Retina Events and DVS Events under a controlled evaluation setting, using the same source videos, network architecture \cite{scheerlinck_fast_2020}, data augmentation and training procedure. We evaluate two complementary tasks: video intensity reconstruction, which tests how much intensity information is preserved, and video saliency prediction, which tries to predict the saliency map derived from human eye-fixation data. Our results show that Retina Events are less suitable for reconstructing absolute intensity frames, indicating that the model discards intensity information through its filtering stages. However, this loss of photometric detail appears to act as an information distillation: on video saliency prediction, Retina Events achieve a 13\% lower validation loss than DVS Events while using the same preprocessing, network, and training setup.

The key contributions of this work are:

\begin{itemize}
    \item A complete mapping of a multi-stage silicon retina model onto the SCAMP-5 pixel-processor array, incorporating biologically motivated functions such as center-surround filtering and spike-based encoding. To the best of our knowledge, this is the first implementation of a multi-stage silicon retina model with center-surround filtering, contrast gain control, and spike-based ON/OFF output on a Pixel Processor Array.

    \item An implementation strategy to mitigate SCAMP-5 constraints, using a superpixel memory architecture to increase memory depth and a 5-bit fixed-point multiplication scheme using analog operations. 

    \item A novel, accessible simulation framework combining an off-the-shelf high-speed sensor (VD55G0) with CUDA acceleration. This allows for the simulation of silicon retina dynamics on standard hardware, enabling easier, further research at the cost of reduced power efficiency in comparison to PPA solution.
    
    \item An empirical evaluation of biologically inspired event representations on downstream reconstruction and saliency-prediction tasks. We show that DVS outperforms silicon retina on the video reconstruction task, however, on video saliency prediction, silicon retina events achieve a loss 13 \% lower than standard DVS events while being 47\% sparser. Code, models, implementation details, and demonstration videos are released on github : \href{https://github.com/mlewandowski0/silicon-retina}{https://github.com/mlewandowski0/silicon-retina}.
    
\end{itemize}


\begin{figure}[htbp]
  \centering
  \includegraphics[width=1.0\textwidth]{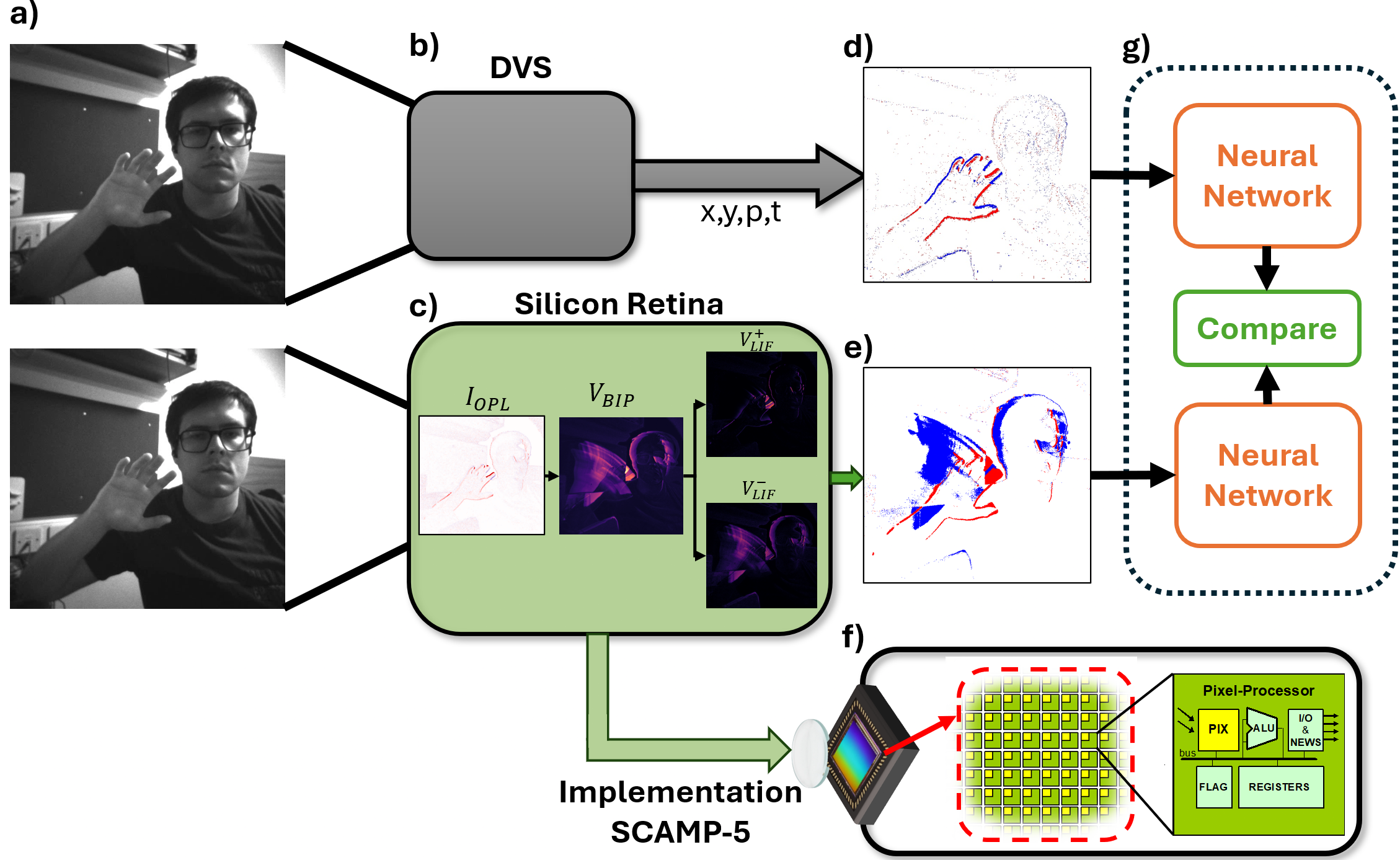}
  \caption{Visual summary of the proposed framework. (a) Standard intensity frames captured by a digital camera. (b) The Dynamic Vision Sensor (DVS) approach, which generates a stream of temporal events $(x,y,p,t)$. (c) The bio-inspired Silicon Retina model investigated in this work. (d) Simulated output of a conventional DVS (temporal contrast only) using the method from~\cite{gehrig_video_2020}. (e) Simulated output from the proposed Silicon Retina model. (f) The SCAMP-5 Pixel Processor Array (analog hardware) utilized to implement the silicon retina directly at the focal plane. (g) The downstream evaluation pipeline (Section~\ref{sec:retina_benefits}), which investigates the benefits of the retinal events for semantic computer vision tasks using large-scale synthetic data.}
  \label{fig:visual_abstract}
\end{figure}

\newpage
\section{Silicon Retina Model}
\label{section:model}

Existing retinal models range from highly detailed biophysical simulations to efficient functional approximations \cite{field_information_2007, baccus_fast_2002,guo_understanding_2014}. Biophysical models often capture non-trivial connectivity, cellular operations and synaptic interactions at the cost of high computational complexity. On the other side, functional models typically use a sequence of filtering stages to efficiently reproduce experimentally observed outputs. Many functional approaches follow a Linear-Nonlinear (LN) architecture, consisting of linear filtering of the visual stimulus, a static nonlinearity, and a spike-generation stage.

In this work, we implement the biologically plausible retina model proposed in \cite{wohrer_virtual_2009}. This functional model extends standard LN architectures by incorporating a biologically plausible Contrast Gain Control mechanism. While it abstracts away specific cellular details and non-trivial topology, it captures essential retinal dynamics - particularly adaptation to varying light conditions - better than static LN models \cite{wohrer_virtual_2009}. We selected this model because it balances biological fidelity with computational efficiency, making it suitable for acceleration on hardware such as FPGAs \cite{philip_neuromorphic_2025}, analog implementations \cite{Prince_tau_cell_2025} or, as in this work, Pixel Processor Arrays.

\newpage
The model implements three sequential stages : 
\begin{enumerate}
    \item \textbf{Outer Plexiform Layer (OPL):} Models photoreceptors and horizontal cells using spatio-temporal filtering to extract edges and motion (center-surround dynamics).
    \item \textbf{Bipolar Cell Layer:} Implements a non-linear feedback mechanism to achieve local contrast gain control.
    \item \textbf{Inner Plexiform Layer (IPL) and Ganglion Cells:} Converts the analog currents into discrete spike events using Leaky Integrate-and-Fire (LIF) neurons.
\end{enumerate}

\subsection{Outer Plexiform Layer (OPL)}
The OPL mimics the synaptic interactions between photoreceptors, horizontal cells, and bipolar cells \cite{wohrer_virtual_2009}. It transforms the input luminance $L$ into a center-surround signal by computing the difference between a narrow (pixel) "center" current $I_C$ and a wider (neighborhood) "surround" current, $I_S$.
Functionally, this stage behaves as a spatio-temporal band-pass filter: spatially, it acts as a Difference of Gaussians (DoG) to detect edges; temporally, the delay between center and surround integration allows the system to respond to motion. The dynamics are modelled by the following equations:
\begin{equation}
 I_\text{C}=T_\text{C}* E_\text{C}*G_\text{C}*L
 \label{c1}
\end{equation}
\begin{equation}
  I_\text{S}=E_\text{S}*G_\text{S}*I_\text{C}
  \label{s1}
\end{equation}
 \begin{equation}
 I_\text{OPL}= \lambda(I_\text{C} - \omega_{OPL} I_\text{S}) 
 \label{opleq}
 \end{equation}
  where $L$ is the input luminance, while $G_\text{C}$ and $G_\text{S}$ denote spatial (Gaussian) filters applied to the center and surround areas, respectively. $T_\text{C}$ denotes a high-pass temporal filter, while $E_\text{C}$ and $E_\text{S}$ represent low-pass temporal filters. The asterisk symbol (*) represents the convolution operation. $\lambda$ is a scalar signifying amplification and $\omega_{OPL}$ denotes how much surround is removed from the center signal.  
\subsection{Contrast Gain Control in Bipolar Cells }
\label{sec:adaptation}
Biological vision must operate across light intensities spanning approximately 14 orders of magnitude. Photoreceptors alone cannot cover this range, so the visual system must adapt continuously to changes in illumination. This adaptation is achieved through multiple gain-control mechanisms, including iris constriction, photoreceptor adaptation, and cortical lateral inhibition \cite{wohrer_virtual_2009,shapley_effect_1978,baccus_fast_2002}. In the retina, this adaptive regulation is implemented partly through contrast gain control between bipolar and amacrine cells, whereby local stimulus contrast dynamically adjusts the gain of the bipolar-cell response. Experimental evidence for this mechanism was obtained by presenting stimuli with different contrasts across a cell’s receptive field and measuring the resulting input-output relationship \cite{wohrer_virtual_2009}.

This mechanism is modeled as a non-linear feedback loop. The bipolar cell membrane potential $V_{BIP}$ is governed by a leaky integration, with the leak conductance $g_{A}$ varying per pixel. The following equation captures this: 
\begin{equation}
     I_{BIP}=C\frac{dV_{BIP}}{dt} = I_{OPL} - g_{A}V_{BIP}
     \label{cvbip}
 \end{equation}
Here $I_{OPL}$ acts as the driving current. The variable conductance $g_{A}$ acts as a divisive feedback term, effectively changing the input gain.  The conductance $g_{A}$ is derived from a spatially ($G_{A}$) and temporally ($E_{A}$) filtered state of the cell, passed through a non-linear function $Q$.
\begin{equation}
       g_{A} = G_{A}*E_{A}*Q(V_{BIP})
       \label{ga}
 \end{equation}
The function $Q(V_{BIP})$ defines how the system adapts to contrast. It is modelled as a parabolic function: 
\begin{equation}
    Q(V_{BIP}) = g^0_{A} + \lambda_{A} V_{BIP}^{2}
     \label{qfunc}
 \end{equation}
In regions of high contrast (large $V_{BIP}$), function output $Q(V_{BIP})$ increases, causing $g_{A}$ to rise. A larger $g_{A}$ increases the "leak" in equation \ref{cvbip}, which reduces the gain and prevents saturation. For low contrast, $g_{A}$ remains near its resting value $g^0_{A}$ which maintains high sensitivity. The parameter $\lambda_{A}$ controls the strength of this feedback loop.

\subsection{Inner Plexiform Layer and Ganglion cells }

The final stage converts the analog bipolar signals into a stream of spikes by modeling the output of the Retinal Ganglion Cells. This is achieved by sending current from non-spiking bipolar and amacrine cells from the previous stage to a spiking ganglion cell model. This behaviour is captured by:
\begin{equation}
     I_{gang} = N(T_2 * V_{BIP})
    \label{eq:I_gang}
 \end{equation}
where $T_2$ is a temporal filter and $N(\cdot)$ is a rectification function used by Wohrer et al \cite{wohrer_virtual_2009}, and models the nonlinear synaptic transmission from bipolar to ganglion cells.

\begin{equation}
    N(V) =
    \begin{cases}
        \frac{i_{G}^{0}}{1 - \lambda_{G}(V-v_{G}^0)/i^{0}_{G}}  & \text{if }V < v_{G}^0 \\
        i_{G}^0 + \lambda_{G}(V-v_{G}^0) & \text{if } V > v_{G}^{0}
    \end{cases}
    \label{eq:nonlinearity}
\end{equation}
where the parameters from equation \ref{eq:nonlinearity} ($i_{G}^0, \lambda_{G}, v_{G}^{0}$) enable modelling of the specific ganglion cell responses \cite{wohrer_virtual_2009}, and change depending on what subtype of ganglion cell is modelled.

The current drives a Leaky Integrate-and-Fire (LIF) neuron, as follows: 
\begin{equation}
    \tau_{m} \frac{dV_{LIF}}{dt} = - (V_{LIF} - V_{rest}) + R \cdot I_{gang}
    \label{eq:Gang_LIF}
\end{equation}

The spiking and reset dynamics are defined as :\begin{align}
\label{eq:spike}
\mathrm{spike}(t) &= 
\begin{cases}
1, & V_{LIF}(t) \ge V_{\mathrm{th}},\\
0, & V_{LIF}(t) < V_{\mathrm{th}},
\end{cases}\\[1ex]
V_{LIF}(t^+) &=
\begin{cases}
V_{LIF}(t) -  V_{\mathrm{reset}}, & V_{LIF}(t) \ge V_{\mathrm{th}},\\
V_{LIF}(t), & V_{LIF}(t) < V_{\mathrm{th}}.
\end{cases}
\label{eq:reset}
\end{align}
If the threshold $V_{th}$ is not reached, $\text{spike}(t) = 0$. This mechanism produces an asynchronous event stream similar to that of a DVS, but derived from a richer, spatio-temporal dynamics of the retinal layer.

In this section, we described a model producing a single event stream. However, standard event-based processing relies on complementary ON and OFF polarity events. The extension to dual-polarity output involves a hardware-specific approximation and is described in Section~\ref{sec:off_events}.

\begin{figure}
  \centering
  \includegraphics[width=1.0\textwidth]{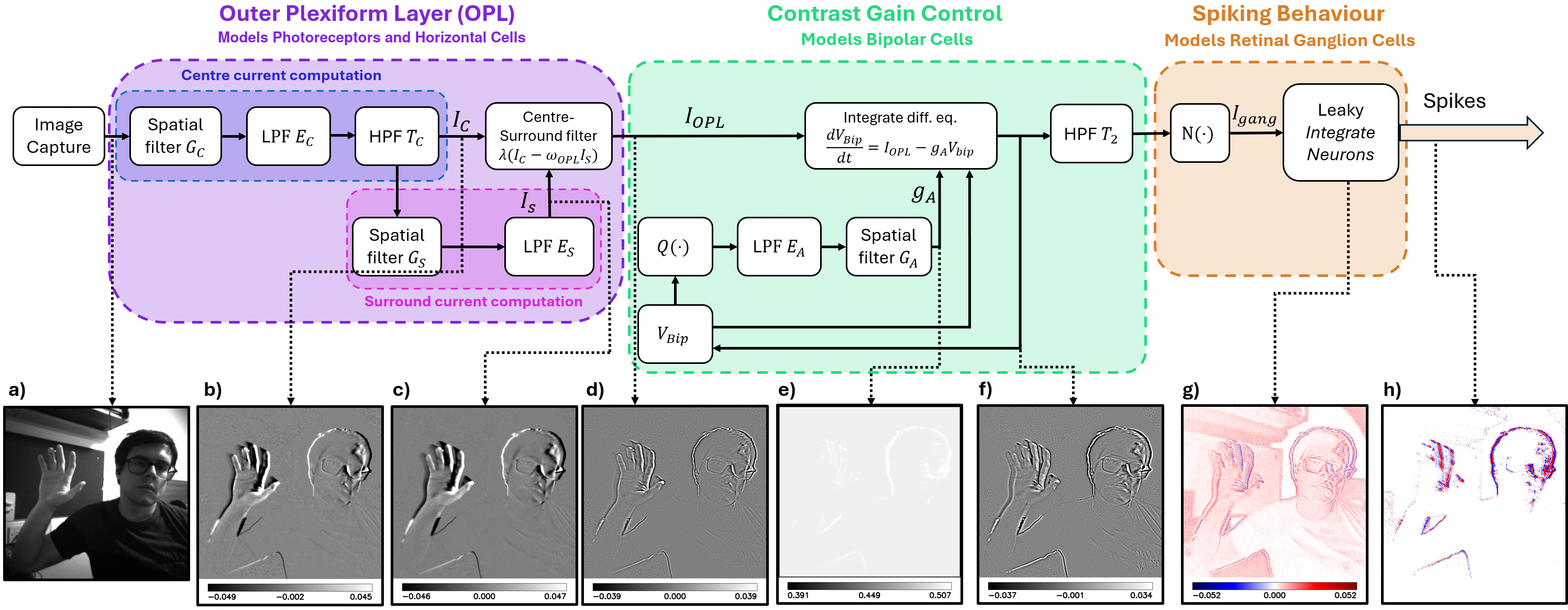}
  \caption{Visualisation of the multi-stage retinal processing pipeline. (a) Input stimulus: Frame from a 100 FPS recording captured via the VD55G0 sensor ($640 \times 600$ pixels). (b) Center current ($I_C$) and (c) Surround current ($I_S$) computed in the Outer Plexiform Layer. (d) The resulting OPL difference current ($I_{OPL}$), highlights edges and movements. (e) The Bipolar Cell membrane potential ($V_{BIP}$), which is modulated by (f) the adaptive local gain ($g_A$). (g) The membrane potential of the Leaky Integrate-and-Fire (LIF) ganglion layer. (h) The final asynchronous spike output generated by complementary ON/OFF LIF neurons ($V_{LIF}^+$ and $V_{LIF}^-$) provides a sparse, event-based representation of the scene.}
  \label{fig:biological_model}
\end{figure}

The model, as well as some of the outputs from the described layers, can be seen in Fig. \ref{fig:biological_model}.

\subsubsection{Generating ON/OFF Polarity Events in Retina Model}
\label{sec:off_events}
Standard event-based algorithms and hardware rely on signed polarity updates (ON and OFF events). In the model described so far, we only generate a single polarity stream of events. To support this, and to mirror the parallel ON/OFF streams, we extend the Ganglion layer to two complementary Leaky-Integrate-Fire neuron channels. These channels are driven by rectified versions of the bipolar output, effectively splitting the signal into positive and negative signals. More particularly, we modify the equation \ref{eq:I_gang} and define the separate driving currents for the ON and OFF channels as : 
\begin{align}
    I_{gang}^{+}  &= N'(+1,T_{2}*V_{BIP}) = N(T_2 * V_{BIP}) \\
    I_{gang}^{-}  &= N'(-1, T_{2} * V_{BIP}) = N(-T_2 * V_{BIP})
\end{align} 
These currents drive two independent LIF neuron populations : 
\begin{align}
    \tau_{m}^+ \frac{dV_{LIF}^{+}}{dt} &= -(V_{LIF}^{+} - V_{rest}^+) + R^+ \cdot I_{gang}^{+} \\
    \tau_{m}^- \frac{dV_{LIF}^{-}}{dt} &= -(V_{LIF}^{-} - V_{rest}^-) + R^- \cdot I_{gang}^{-}
\end{align}

This configuration produces two "complementary" event streams (positive and negative polarity), which makes the output compatible with standard event-processing algorithms. In this work, we assume that $\tau_{m}^+ = \tau_{m}^{-}
$, $V_{rest}^{+} =V_{rest}^{-}$ and $R^{+}=R^{-}$ for simplicity; however, this can be changed easily in the SCAMP-5 implementation and digital simulation framework.

\section{Implementation of Silicon Retina Model on Pixel-Processor Array}

\subsection{SCAMP-5 Pixel Processor Array}
\label{sec:scamp5_hardware}
In this section, we describe briefly the capabilities of the Pixel Processor Array used in this work. Instead of streaming full raw frames off-chip to an external host for processing, PPA performs massively parallel computation at the point where light is captured.  The chip has a $256 \times 256$ grid of heterogeneous Processing Elements (PEs), where each PE corresponds to a single pixel ( Fig. \ref{fig:visual_abstract} d) ).

Each PE is an analog "microprocessor" containing:
\begin{enumerate}
    \item a photodiode for light acquisition,
    \item an analog Arithmetic Logic Unit (ALU) capable of basic arithmetic (addition, subtraction, division by 2) and logic operations.
    \item six general-purpose analog registers (labelled A through F) for storing continuous variables (which can be discretized to 8-bit signed integers).
    \item seven single-bit, volatile digital registers (DREGs) for local logic, and a few more for conditional execution.
    \item a local interconnect network allowing data exchange with the four immediate neighbors (North, East, South, West).
\end{enumerate}

By operating entirely within the pixel array, SCAMP-5 can compute at kilohertz framerates while consuming just under 1 watt of power. This efficiency has enabled applications and algorithms infeasible with conventional camera sensors, including gaze tracking \cite{bose_pixel_2022} at 10,000 frames per second (FPS), low-bandwidth feature descriptors and tracking \cite{chen_feature_nodate, bose_descriptor--pixel_nodate} at a few thousand FPS, and real-time passive depth estimation from depth-from-focus \cite{martel_real-time_2018}. Despite hardware constraints, it was also possible to run small neural networks \cite{so_pixelrnn_2023}, producing compact latent representations for transmission. 

While SCAMP-5 is not an event camera in the conventional sense, as it does not contain asynchronous, hard-wired intensity-change detection circuits, it can produce event-like outputs with greater flexibility. For example, values stored in analog registers can be thresholded to trigger sparse readouts, effectively generating programmable events. In contrast to a DVS, whose event-generation circuitry is dedicated to detecting logarithmic brightness changes, SCAMP-5 allows the event-generation criterion to be defined by the programmed pixel-level computation. This makes it possible to emit sparse events from more complex analog processing pipelines, including the retinal model proposed in this work. 
However, it is important to note that the SCAMP-5 photodiodes operate with a linear response, and thus do not inherently possess the high dynamic range (HDR) characteristic of logarithmic DVS pixels.

\subsection{Pixel Processor Array Programming and Constraints}
\label{sec:constraints}
Implementing a retinal model on SCAMP-5 requires more than a direct translation of the equations in Section~\ref{section:model}. Although the architecture enables massively parallel focal-plane computation, its Single Instruction, Multiple Data (SIMD) execution model, local communication, limited analog memory, and noisy arithmetic impose constraints that strongly influence the algorithm design. The main constraints are:

\begin{enumerate}[itemsep=4pt, parsep=0pt, topsep=4pt]
    \item \textbf{Analog Arithmetic and Noise:} majority of the model computation is performed in the analog domain. Unlike digital logic, these operations are subject to noise and variability. Furthermore, precision is effectively limited to approximately 8 bits, and errors accumulate with each sequential operation. While it is possible to develop well-working algorithms, this becomes problematic for DSP-heavy algorithms, which require many multiplication/division operations.
    \item \textbf{Volatile Analog Storage:} The six analog registers per PE rely on capacitors to store charge. Consequently, the data is volatile and subject to change over time. While this decay is negligible over micro- or millisecond timescales, it becomes significant over longer timescales, posing a challenge for algorithms that require persistent state variables, such as temporal filters or differential equations used in retinal modelling.
    \item \textbf{Neighborhood communication only:} SCAMP-5 follows a SIMD programming model, which means that the same instruction is broadcast to all pixels simultaneously. Data exchange is local: a pixel can communicate directly only with its four immediate neighbors. While some global operations are supported (e.g. summing all pixels), efficient implementations rely on neighbor-to-neighbor interactions.
    \item \textbf{I/O bottleneck} : while the internal array communication is very fast (due to massive parallelism), the interface to the external world is relatively slow. Offloading full frames at 8 bit resolution is limited to approximately 60 FPS. To achieve efficient algorithms and maintain kilohertz operating speeds, algorithms must process data internally and transmit only sparse results to the host.
    \item \textbf{MCU-PPA Co-design:} The system architecture of the PPA includes a standard microcontroller (MCU) acting as a controller for the parallel PPA accelerator. The programmer writes a single C++ application where standard logic runs on the MCU, while SIMD kernels are dispatched to the PPA. This offers a unique opportunity to offload scalar operations to the MCU while performing parallel operations on the PPA.
    
\end{enumerate}

In this section, we describe the implementation of the silicon retina model (Section \ref{section:model}) under these constraints. More particularly, we provide details on how we addressed the limited memory via a superpixel architecture and approximated the required dynamics using analog primitives. This allowed for a low-power, analog, high-framerate simulation of the retina.

\subsection{SCAMP-5 Implementation}
\label{section:implementation_strategy}

To simplify the implementation, we consider only a single class of bipolar cells ( $V_{BIP}^+$ = $V_{BIP}^-$ \cite{philip_neuromorphic_2025}) and assume symmetric processing of surround signals.  The complete computational pipeline is outlined in Algorithm \ref{alg:top_silicon_retina}. 

\begin{algorithm}
\caption{Silicon Retina Model on SCAMP-5}
\label{alg:top_silicon_retina}

\KwIn{Current image $L^t$; previous states $E_C^{t-1}$, $T_C^{t-1}$, $E_S^{t-1}$, $V_{\mathrm{BIP}}^{t-1}$, $E_A^{t-1}$, $T_2^{t-1}$, $V_{\mathrm{LIF}}^{+,t-1}$, $V_{\mathrm{LIF}}^{-,t-1}$}
\KwOut{ON/OFF event maps $S^{+,t}$, $S^{-,t}$; updated states at time $t$}

\textbf{1. Outer Plexiform Layer (OPL)};

$G_C^t \leftarrow \mathrm{blur}(L^t)$ \tcp*{Spatial smoothing}
$E_C^t \leftarrow \mathrm{low\_pass\_filter}(E_C^{t-1}, G_C^t)$ \tcp*{Temporal smoothing}
$T_C^t \leftarrow \mathrm{high\_pass\_filter}(T_C^{t-1}, E_C^t)$ \tcp*{Temporal high-pass response}
$G_S^t \leftarrow \mathrm{blur}(T_C^t)$ \tcp*{Surround pathway}
$E_S^t \leftarrow \mathrm{low\_pass\_filter}(E_S^{t-1}, G_S^t)$; \\
$I_{\mathrm{OPL}}^t \leftarrow \lambda \left(T_C^t - \omega_{\mathrm{OPL}} E_S^t\right)$ \tcp*{Center-surround current}

\textbf{2. Contrast Gain Control in Bipolar Cells};

$E_A^t \leftarrow \mathrm{low\_pass\_filter}\left(E_A^{t-1}, Q(V_{\mathrm{BIP}}^{t-1})\right)$ \tcp*{Feedback state}
$g_A^t \leftarrow \mathrm{blur}(E_A^t)$ \tcp*{Local gain conductance}
$V_{\mathrm{BIP}}^t \leftarrow V_{\mathrm{BIP}}^{t-1} + \Delta t \left(I_{\mathrm{OPL}}^t - g_A^t V_{\mathrm{BIP}}^{t-1}\right)$ \tcp*{Euler integration}

\textbf{3. Ganglion Cells and ON/OFF Spike Generation};

$T_2^t \leftarrow \mathrm{high\_pass\_filter}(T_2^{t-1}, V_{\mathrm{BIP}}^t)$ \tcp*{Bipolar temporal response}

\For{$p \in \{+1,-1\}$}{
$I_{\mathrm{gang}}^{p,t} \leftarrow N'\left(p,  T_2^t\right)$ \tcp*{ON for $p=+1$, OFF for $p=-1$}

$V_{\mathrm{LIF}}^{p,t} \leftarrow V_{\mathrm{LIF}}^{p,t-1}
+ \frac{\Delta t}{\tau_m}
\left(R \cdot I_{\mathrm{gang}}^{p,t}
- \left(V_{\mathrm{LIF}}^{p,t-1} - V_{\mathrm{rest}}\right)\right)$ \tcp*{Euler integration}

\If{$V_{\mathrm{LIF}}^{p,t} \geq V_{\mathrm{th}}$}{
    $V_{\mathrm{LIF}}^{p,t} \leftarrow V_{\mathrm{LIF}}^{p,t} - V_{\mathrm{reset}}$\;
    $S^{p,t} \leftarrow 1$ \tcp*{Spike generated}
}
\Else{
    $S^{p,t} \leftarrow 0$\;
}

}

\end{algorithm}

\subsubsection*{Implementation Challenges}
Directly mapping Algorithm \ref{alg:top_silicon_retina} to the SCAMP-5 PPA is challenging.  As detailed in section \ref{sec:constraints}, each PE contains only six analog registers. A naive count of the state variables required by the filters and differential equations indicates at least 7 persistent (frame-to-frame) registers per pixel. This exceeds the physical capacity of a single PE.

Furthermore, the algorithm relies heavily on global scalar multiplication ($\lambda, \omega, R, \frac{\Delta t}{\tau_m}$) and per-pixel multiplication (every PE has its own $g_{A}$). Since SCAMP-5 lacks a hardware multiplier, these operations must be approximated using the limited ALU instruction set (addition, subtraction, division-by-two, and binary operations). Finally, the deep pipeline of sequential operations makes the system susceptible to noise accumulation, particularly in the non-trivial contrast gain control layer.

To resolve the lack of memory and enable complex arithmetic, we introduce a superpixel memory layout and a fixed-point analog approximation scheme, described below.

\subsection{Superpixel memory architecture and multiplication approximation}
\label{section:superpixel_arithmetic}
\subsubsection{Memory architecture}
\begin{figure}[htbp]
  \centering
  \includegraphics[width=0.6\textwidth]{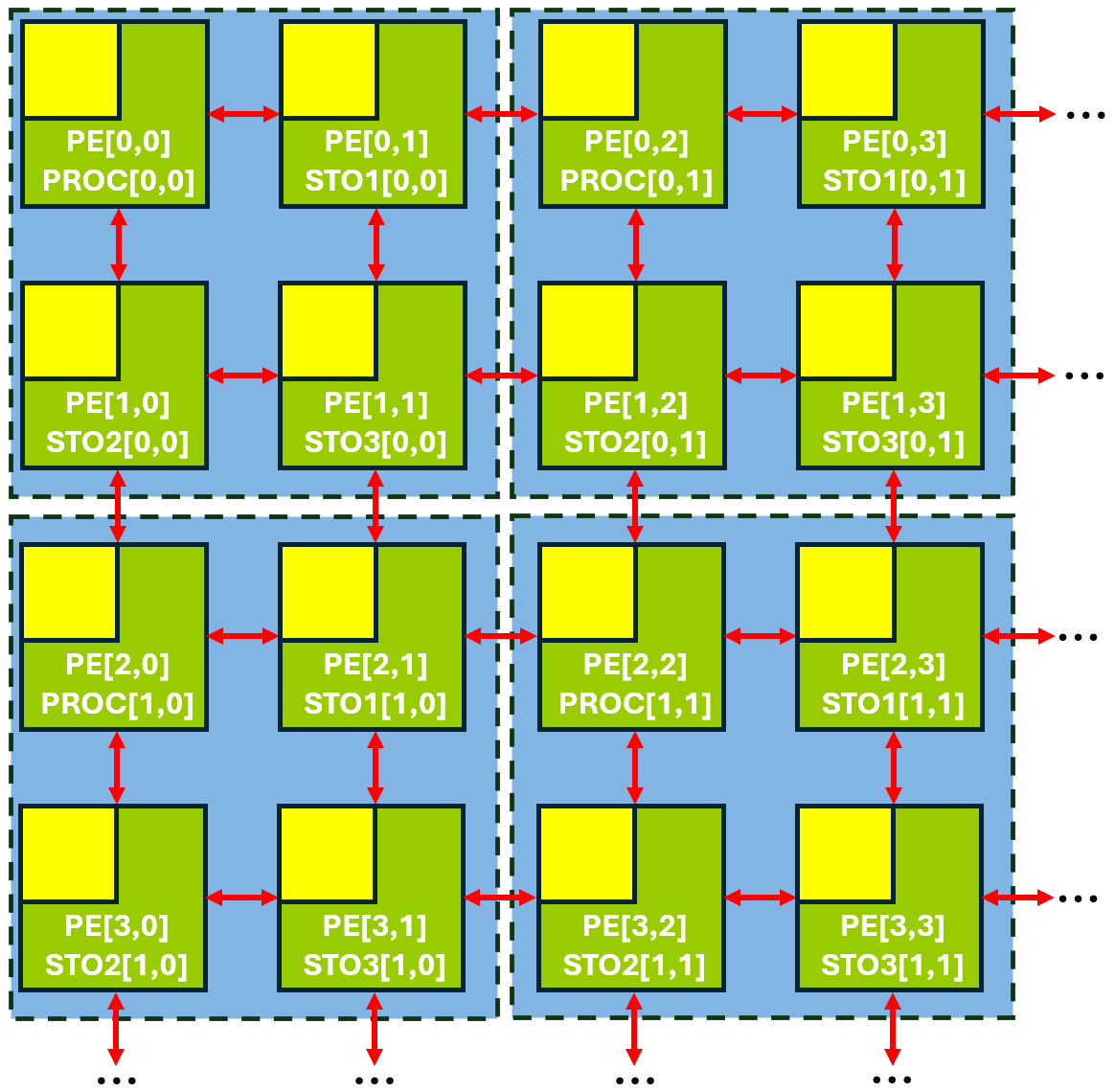}
\caption{Superpixel architecture on SCAMP-5. The physical array is partitioned into $2 \times 2$ blocks. The top-left PE(\texttt{PROC}) performs computation, while the other three (\texttt{STO}) act as extended analog memory. This almost quadruples the available register count per effective pixel, enabling the computation of retina dynamics.}  \label{fig:SCAMP5_superpixel}
\end{figure}

To increase the memory per pixel, we logically partition the $256 \times 256$ physical array into $2 \times 2$ blocks, creating a $128 \times 128$ array of "superpixels" (Fig. \ref{fig:SCAMP5_superpixel}). 

In this configuration, the four physical PEs within a block act as a single logical unit with shared resources. One PE is designated as the processor (\texttt{PROC}), responsible for active computation and image capturing. The remaining three PEs serve as Storage Units (\texttt{STO1}, \texttt{STO2}, \texttt{STO3}), effectively increasing the available analog memory from 6 to 24 registers per logical pixel. This allows us to persist the multiple filter states ($E_C, T_C, E_S$) and membrane voltage values ($V_{BIP}$, $V_{LIF}$), required by Algorithm \ref{alg:top_silicon_retina}, at the cost of halving the spatial resolution. 

Data transfer between \texttt{PROC} and \texttt{STO} units is handled via local North/East/West/South (NEWS) communication instructions. Spatial operations (e.g. Gaussian blur via the resistive network) require more consideration for these superpixels: we replicate the value of the active register (for example, for $E_{A}^{t-1}$) from the \texttt{PROC} unit to all three \texttt{STO} units prior to the spatial operation. This ensures that the hardware's resistive grid operates correctly across the active register plane.

\subsubsection{Fixed-point analog multiplication}
\label{sec:fixed_point}
The retinal model requires two distinct forms of multiplication:
\begin{enumerate}
    \item \textbf{Global Scaling:} Multiplying an image by a global constant $\alpha$ (e.g., resistance $R$)
    \item \textbf{Local modulation:} Multiplying an image by a spatially varying map (e.g, gain $g_A$).
\end{enumerate}
Since SCAMP-5 lacks a hardware multiplier, we approximate these operations using 5-bit fixed-point arithmetic. We define a scalar multiplier $\alpha$ as a binary fraction:
\begin{equation}
    \alpha \approx \sum_{i=0}^4 b_i\cdot2^{-i} = b_{0} + \frac{b_1}{2} + \frac{b_2}{4} + \frac{b_3}{8} + \frac{b_4}{16}
\end{equation}

where $b_i \in \{0,1\}$. The product $A' = \alpha \cdot A$ is computed by accumulating progressively halved versions of the register $A$:
\begin{equation}
    A' = \sum_{i=0}^{4} b_{i} \left( \frac{A}{2^i} \right)
    \label{eq:scamp_mult}
\end{equation}
In hardware, the division by powers of 2 is implemented efficiently using the SCAMP-5 \texttt{diva} instruction (division by 2) sequentially. The binary coefficients $b_i$ gate the addition of these partial terms.

For global constants, the coefficients $b_i$ are stored in the microcontroller. For spatially varying maps, like the gain $g_A$, the coefficients are firstly digitized into four single-bit planes (where the most significant bit is always zero) stored in the local digital registers (DREGs). These local bits then serve as the coefficients $b_i$ for each pixel, enabling parallel multiplication with pixel-wise multiplicand.

For amplification factors where $\alpha > 1$ (e.g., gain $\lambda$), we utilize repeated addition ($A \leftarrow A + A$) to perform multiplication by 2 prior to the accumulation of fractions.

\subsubsection*{Tradeoff in Precision vs Noise}

Despite its simplicity and compatibility with the SCAMP-5 instruction set, the fixed-point multiplication scheme introduces significant amount of noise due to repeated analog operations. This degradation of signal is particularly visible when input signals are weak (e.g. at a very high frame rate) or when the multiplier $\alpha$ is small (requiring many division operations). Empirical analysis indicated that a 5-bit representation offers the optimal balance for this architecture. Lower bitwidths fail to capture the necessary range, while higher bitwidths result in too much degradation of the signal-to-noise ratio.

\subsection{Mapping the algorithm to SCAMP-5}
\label{section:mapping}
Leveraging the superpixel memory architecture and the fixed-point arithmetic scheme, we map the functional components of the retina model (Algorithm \ref{alg:top_silicon_retina}) to the hardware. Below, we detail the specific SCAMP-5-compatible approximations used for every step of the algorithm. 

\begin{enumerate}
    \item \textbf{Low-Pass Temporal Filters (IIR Approximation):}
    We implement low-pass filtering using a discretized Infinite Impulse Response (IIR) structure:
    \begin{equation}
        A[t] \leftarrow A[t-1] + \alpha (B[t] - A[t-1])
        \label{eq:lpf}
    \end{equation}
    Where register $A$ holds the persistent filter state and $B$ is the current input. The decay coefficient $\alpha$ is applied using the fixed-point multiplication strategy described in equation \ref{eq:scamp_mult}.
    \item \textbf{High-Pass Temporal Filters (Differencing):}
    A standard IIR high-pass response is derived by computing an IIR low-pass state and subtracting it from the input signal: 
\begin{equation}
    A_{HPF2}[t] = B[t] - \big( A_{LPF}[t-1] + \alpha (B[t] - A_{LPF}[t-1]) \big) = (1-\alpha)( B[t] - A_{LPF}[t-1])
    \label{eq:standard_high_pass}
\end{equation}
    This approach proved problematic : the OPL stage required to compute two filters (LPF and HPF) sequentially, which produced excessively noisy output. We therefore use a simpler frame-differencing scheme that yields qualitatively better results. Given an input $B$ and output $A$:
    \begin{equation}
        A[t] \leftarrow B[t] - B[t-1]
        \label{eq:hpf}
    \end{equation}
    Although this is a large simplification of the model, it minimizes the accumulation of analog noise across time steps, providing more stable behaviour.
    \item \textbf{Spatial Blur (Resistive Grid):}
    SCAMP-5 hardware natively supports spatial smoothing via a resistive network. We utilize this to implement the Gaussian kernels ($G_C, G_S, G_A$ efficiently). This operation is asynchronous and introduces negligible noise \cite{carey_100000_2013}.

    \item \textbf{Feedback Nonlinearity ($Q$-function):}
    The biological model specifies a quadratic dependence on the bipolar potential ($V_{BIP}^2$). While SCAMP-5 does have a squarer circuit, it has relatively large spatial non-uniformity. We approximate this non-linearity using an absolute value function : 
    \begin{equation}
        Q(V_{BIP}) \approx g_{A}^0 + \lambda_A' \cdot |V_{BIP}|
        \label{eq:feedback_relaxed}
    \end{equation}
    This linear rectification simplifies the contrast gain control mechanism. In Section \ref{sec:retina_benefits}, we demonstrate that replacing the quadratic form with this absolute value approximation does not negatively impact performance on video saliency tasks.

    \item \textbf{Differential Equations (Euler Integration):}
    We solve the ordinary differential equations governing the Bipolar and Ganglion layers using an explicit forward Euler method. The update rule for the bipolar membrane potential is implemented as :
    \begin{equation}
        V_{BIP}[t] \leftarrow V_{BIP}[t-1] + p_{1} \cdot (I_{OPL} - g_{A} \cdot V_{BIP}[t-1])
    \end{equation}
    Similarly, the Leaky Integrate-and-Fire (LIF) neuron state is updated as:
    \begin{equation}
        V_\text{LIF}[t] \leftarrow V_\text{LIF}[t-1] + p_2 \cdot I_\text{gang} -  p_3 \cdot |V_\text{LIF}[t-1] - V_\text{rest}|
        \label{eq:LIF_int}
    \end{equation}
    Here, the parameters $p_1, p_2, p_3$ absorb the time step $\Delta t$ and capacitance $C$, and are applied using fixed-point multiplication.
    \item \textbf{Nonlinearity \( N(\cdot) \)}:  
    The complex function \( N(\cdot) \), defined in \cite{wohrer_virtual_2009, philip_neuromorphic_2025}, was approximated using a simple rectification:
    \begin{equation}
        N(x) \approx \max(0, x) + I_\text{bias}
    \end{equation}
    This approximation roughly preserves the shape of the original function while remaining implementable with basic SCAMP-5 primitives.
\end{enumerate}

\subsubsection{Performance}
The performance of the implemented system is determined by the I/O and lighting conditions. In a purely internal model, where processing occurs without external readout, the retina model executes at approximately 10,000 FPS. However, when streaming binary spike frames off-chip (effectively single-bit images), the frame rate is bottlenecked by the digital interface to approximately 580 FPS for single channel of events. For a pair of 2 LIF neurons (ON/OFF), as described in algorithm \ref{alg:top_silicon_retina} we can achieve at most 280 FPS. In practical scenarios (indoor ambient light without artificial lighting), the system performs reasonably well in the 100-200 FPS range, with a sub-watt power consumption. Higher framerates are possible, however at very high framerates the photodiode integration time decreases, producing weaker input signals. When processed through the multi-stage analog pipeline, these weak signals are more susceptible to arithmetic noise accumulation which degrades the final output quality.  
Rather than streaming full binary spike maps, SCAMP-5 can transmit only the coordinates of active (spiking) pixels, reducing output bandwidth in sparse scenes. However the effective framerate  of this event-based readout is highly dependent on scene activity, lighting and motion. Furthermore, the SCAMP-5 event output is much slower than from standard DVS cameras.

\begin{figure}
  \centering
   \includegraphics[width=1.0\textwidth]{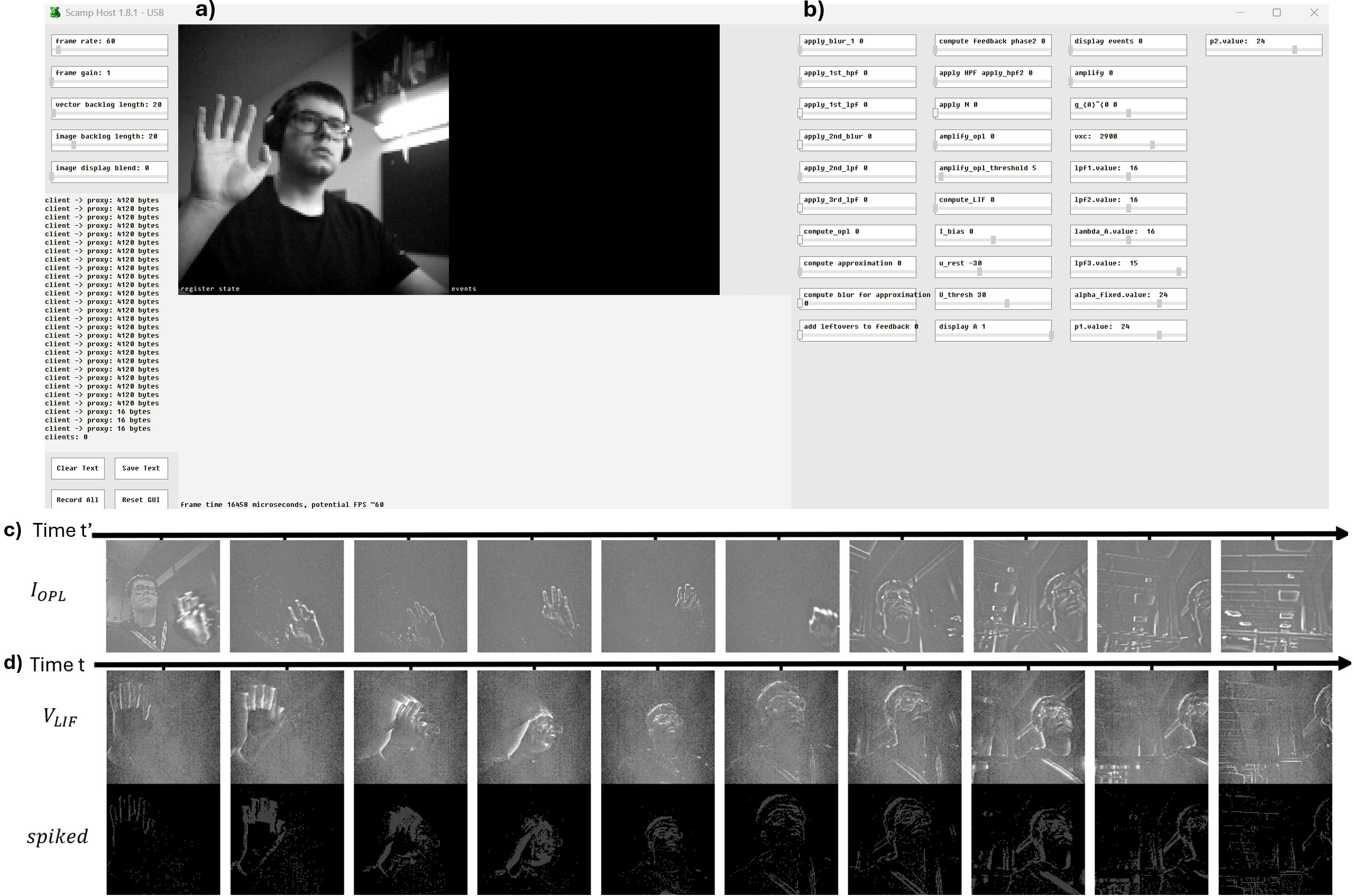}
  \caption{Overview of the SCAMP-5 silicon retina software interface. a) and b) show the host control GUI  used to configure and monitor the retina model in real-time. b) Detailed view of the parameter configuration panel : filter coefficients are adjustable as fixed-point integers (e.g., Low Pass Filter coefficients) and individual pipeline stages can be toggled for debugging. Due to readout bandwidth and memory constraints, visualisation is limited to a single target register at a time. However, higher frame rates can be achieved by streaming only events using the \texttt{display\_events} flag. (c) Visualisation of the intermediate OPL current ($I_{OPL}$). (d) visualisation of the Ganglion Layer output, showing the membrane potential $V_{LIF}^+$ and the resulting binary spike map ($S^{+}$) for the ON neurons.  }
\label{fig:retina_output_gui}
\end{figure}

The host interface, parameter controls, and examples of
intermediate retinal signals and spike outputs are shown
in Fig.~\ref{fig:retina_output_gui}.\textbf{}

\subsection{Simulation Framework and Digital Capture}
\label{sec:simulation}
While pixel-processor arrays offer unparalleled power efficiency and low latency, their specialized programming model and limited availability can hinder widespread experimentation. To enable broader exploration of silicon retina models, we developed a complementary digital simulation framework, based on off-the-shelf sensor and popular GPU-accelerated framework (CUDA). This tool allows researchers to experiment with retinal dynamics using standard hardware, enabling easier debugging, parameter tuning and large-scale data generation.

One of the fundamental challenges in simulating retinal dynamics is the requirement for high temporal resolution input. Conventional CMOS image sensors typically operate at 30-120 FPS and fail to capture the fast transient dynamics necessary for accurate modeling. We address this limitation through two complementary approaches : direct high-speed capture and neural temporal upsampling.

\subsubsection{Direct Capture via High-Speed CMOS sensor}
To obtain real-world high-speed data without PPA hardware we utilize the VD55G0 sensor, which is a recent low-cost, high-sensitivity global shutter CMOS image sensor. Although it lacks in-sensor processing capabilities, it supports frame rates and resolutions comparable to our SCAMP-5 implementation. Specifically, the sensor can capture $160 \times 120$ frames at 760 FPS or $128 \times128$ frames at 640 FPS.

This performance closely aligns with the performance achieved by SCAMP-5 in this work, which is approximately 580 FPS for binary frames. By streaming these high-speed raw frames to a GPU, we can run the retinal model in software (implemented in CUDA) to produce event streams that closely approximate the output of the physical silicon retina. This provides an accessible alternative to the hardware system for algorithm development.

\subsubsection{Synthetic Data via Neural Upsampling}
\label{ref:neural_upsampling}
To enable scalable training of downstream neural networks, we require large, high framerate videos, beyond what is currently available or can be easily captured. We adopt the video interpolation strategy introduced by Gehrig et al. \cite{gehrig_video_2020}. This method uses a neural network to temporally upsample standard low-framerate videos, generating high-framerate sequences with smooth, sub-pixel optical flow. 

While the original work utilized these interpolated frames to generate standard DVS events, we can repurpose the pipeline to drive our CUDA-based silicon retina model. This allows us to convert existing large-scale video datasets into synthetic "retina events" streams, which allow the training and evaluation of event-based algorithms without the need for new physical recordings.

\subsubsection*{Simulation Implementation and Performance}
We implemented a modular simulation of the retina model using C++ with CUDA acceleration. This software provides a Python interface for easy integration with other frameworks and supports extensive parameterisation, including kernel sizes, different filter approximations and layer-specific constants.

Benchmarks were conducted on a laptop equipped with mid-range, modern NVIDIA GPU (RTX 5070, 70W power limit) using 8-bit grayscale videos of resolution $256\times256$ and containing 300 frames. The simulator allows for three operational scenarios : 
\begin{enumerate}
    \item \textbf{All States Output:} Returns all states of every layer (OPL, Bipolar, Ganglion) for visualization and debugging. This is computed for input video.
    \item \textbf{Events Only:} Returns only the sparse binary spike map for whole input video.
    \item \textbf{Online Mode:} instead of processing whole video, this mode simulates frame-by-frame processing typical of a live camera feed. This mode was intended to use with VD55G0 or other camera, example visualization code is provided.
\end{enumerate}

As detailed in Table \ref{tab:GPU_kernels}, the simulator achieves high throughput. In "Events Only" mode, the kernel executes at over 30,000 FPS, demonstrating that the computational load of this particular retina model is negligible for modern GPUs. Even with full memory transfers overheads, the system sustains over 11,000 FPS, making it highly efficient for generating large-scale synthetic datasets from upsampled video. For real-time applications, the online mode comfortably supports 500-800 FPS output of high-speed sensors like VD55G0, which matches the SCAMP-5 framerates, although the CPU/GPU consumes approximately 80 times more power.

\begin{table}[htbp]
  \centering
  \caption{Performance Benchmarks of the CUDA Retina Simulator ($256 \times 256$ input).}
  \label{tab:GPU_kernels}
  \renewcommand{\arraystretch}{1.2}
  \begin{tabular}{lccc}
    \toprule
    \textbf{Operation Mode} & \textbf{\shortstack{Full State\\Output (FPS)}} & \textbf{\shortstack{Events\\Only (FPS)}} & \textbf{\shortstack{Online\\(Frame-by-Frame)}} \\
    \midrule
    \hspace{3mm} Kernel Compute Only & 14,152 & 30,439 & 6,057 \\
    \hspace{3mm} + Host/Device Transfer & 1,283 & 11,383 & 800\textsuperscript{*} \\
    \bottomrule
  \end{tabular}
  \begin{tablenotes}
    \small
    \item \textsuperscript{*} Includes disk I/O overhead. Otherwise, the images were in CPU memory and transfer is done from CPU to GPU and back.
  \end{tablenotes}
\end{table}

\subsubsection{Comparison to other work}
In this work, we presented two distinct hardware approaches for generating biologically plausible retinal events \cite{wohrer_virtual_2009, philip_neuromorphic_2025}.
The SCAMP-5 PPA implementations offers extreme power efficiency (sub-watt) and low-latency processing directly at the focal plane, at the cost of analog noise and more complex programming model. Conversely, the digital simulation framework (VD55G0 and GPU) allows for greater flexibility and easier development, utilizing standard CUDA C++ and Python at the cost of 2-3 orders of magnitude higher power consumption.

Table \ref{tablecompare} contextualizes our contributions against other silicon retina implementations. Unlike standard temporal contrast sensors (i.e. event-camera, DVS) which lack spatial processing, our implementation enable full spatio-temporal dynamics, including center-surround filtering, contrast gain control and leaky-integrate-fire neurons. Notably, the PPA implementation is the first to achieve this programmability within the analog domain, allowing for software-defined control of retina parameters.

\begin{table*}[!t]
\centering
\caption{Comparison of the proposed systems with existing Silicon Retina implementations.}
\resizebox{\textwidth}{!}{%
\begin{threeparttable}
\rowcolors{2}{gray!10}{white}
\renewcommand{\arraystretch}{1.2}
\begin{tabular}{lcccccccc}
\toprule
\rowcolor{gray!30}
\textbf{Parameter} & \textbf{Lichtsteiner} & \textbf{Costas-Santos} & \textbf{Posch} & \textbf{Leñero-Bardallo} & \textbf{Brandli (DAVIS)} & \textbf{Hayashida} & \textbf{This Work} & \textbf{This Work} \\
\rowcolor{gray!30}
 & \textbf{\cite{lichtsteiner_2008}} & \textbf{\cite{costas-santos_spatial_2007}} & \textbf{\cite{posch_live_2010}} & \textbf{\cite{lenero-bardallo_five-decade_2010}} & \textbf{\cite{brandli_live_2014}} & \textbf{\cite{hayashida_retinal_2017}} & \textbf{(SCAMP-5 PPA)} & \textbf{(Digital Sim)} \\
\midrule
Year & 2006 & 2007 & 2010 & 2010 & 2014 & 2017 & 2026 & 2026 \\\textbf{Functionality} & Temporal & Spatial & Temporal & Spatial & Temporal + APS & Spatio-temp. & \textbf{Spatio-temp.} & \textbf{Spatio-temp.} \\
 Programmable & No & No & No & No & No & Yes (FPGA/HDL) & Yes (C++/SCAMP-5 ISA) & Yes (CUDA C++/Python) \\ 
 Hardware & Analog & Analog & Analog & Analog & Analog & Analog+FPGA & mixed signal & Analog + Digital readout/processing \\ 
 Technology & 0.35$\mu$m & 0.35$\mu$m & 0.18$\mu$m & 0.35$\mu$m & 0.18$\mu$m & —\tnote{a} & 0.18$\mu$m & VD55G0 : 45nm, GPU : 5nm \\ 
 Array size & $128\times128$ & $32\times32$ & $304\times240$ & $32\times32$ & \begin{tabular}[c]{@{}c@{}}$240\times180$ (DVS) \\ $240\times180$ (APS)\end{tabular} & $128\times128$ & \begin{tabular}[c]{@{}c@{}}$256\times256$ (PEs) \\ $128\times128$ (retina)\end{tabular} & \begin{tabular}[c]{@{}c@{}}$644 \times 604$ (185-210FPS) \\ $128\times128$ (max 640 FPS)\end{tabular} \\ 
Pixel Pitch ($\mu$m) & $40\times40$ & $58\times56$ & $30\times30$ & $81.5\times76.5$ & $18.5\times18.5$ & $178\times154$ & $32\times32$ & $2.61 \times 2.61$ \\
\midrule
\textbf{Performance} & & & & & & & & \\
 \begin{tabular}[c]{@{}c@{}} Throughput / Mode \end{tabular} & Event & Event & Event & Event, 22 Meps &  Event, 12-50 Meps & 200 fps  & \begin{tabular}[c]{@{}c@{}} 10,000 FPS (no output) \\ 580 FPS (binary map) \\ 100-200fps (without art. light) \tnote{b}  \end{tabular} & \begin{tabular}[c]{@{}c@{}} $185-210fps \ @\ 644 \times 604$ \tnote{c} \\ $640fps \ @\ 128\times128$ \tnote{c} \\ $430fps \tnote{c} \ @\ 320\times240$  \tnote{c} \end{tabular}  \\ 
Power/Pixel & 400 nW & 9.7 $\mu$W & 1.3 $\mu$W & 264 nW & 347 nW & 21.7 $\mu$W & 18.7 $\mu$W\tnote{d} & High (GPU) \\
Dyn. Range (dB) & >120 & — & >120 & — & >120 & — & $\approx$60 (Linear) & $\approx$60 (Linear) \\
\midrule
\textbf{Biological Features} & & & & & & & & \\
 Spatial Filtering & No & Diffusive grid & No & Diffusive grid & No & Diffusive grid & Diffusive grid & Digital filter \\ 
 Contrast gain control & No & No & No & No & No & No & Yes \tnote{e} & Yes \\

Luminescence adaptation & No & No & No & No & No & No & Yes & Yes \\

Tonic and phasic cells & No & No & No & No & No & No & Yes & Yes \\ \bottomrule
\end{tabular}
\begin{tablenotes}
\small
\item[a] Values not reported in the original publication.
\item[b] Without artificial lighting, measured in office with few windows ( natural lighting). Higher framerates are achievable; however, the output quality degrades
\item[c] Values from datasheet. With the USB-C evaluating development kit it is possible to obtain 200 FPS at 322x302, and its possible to obtain higher framerates using MIPI CSI-2 board
\item[d] Typically 1-2W of total system power when outputting frames which is most expensive operation int terms of power. 
\item[e] Implemented on SCAMP-5, though requires tuning as it is sensitive to ambient lighting levels.
\end{tablenotes}
\end{threeparttable}
}
\label{tablecompare}
\end{table*}

\section{Benchmarking Spatio-Temporal Retina Events}
\label{sec:retina_benefits}

\subsection{Introduction}
Having established an analog-based implementation (SCAMP-5) and a flexible digital simulation (CUDA), we now address the research question: Can the richer spatiotemporal representation of the biological retina model offer advantages over standard DVS representation in downstream computer vision tasks ? 

Standard event cameras primarily encode temporal contrast, whereas the proposed silicon retina model incorporates spatial filtering, temporal dynamics, contrast gain control, and spike generation directly into the event formation process. To evaluate whether these additional mechanisms are useful, we compare the proposed ``Retina Events'' against standard ``DVS Events'' on neural-network-based downstream tasks.

A direct hardware-only evaluation is difficult for this comparison. Although the SCAMP-5 implementation demonstrates that the model can be executed on-sensor, the exploration of usefulness of this model requires paired input sequences, ground-truth task labels, and matched event streams generated from the same visual input. This is particularly challenging for the tasks considered in this work. For intensity reconstruction, a hardware comparison would require synchronized intensity frames, DVS events, and retina events from the same scene. For saliency prediction, it would additionally require human eye-fixation annotations for the recorded sequences. Collecting such large-scale, labeled datasets with both DVS and SCAMP-5 retina outputs would be impractical. We therefore use synthetic data generation to isolate the effect of the event representation itself: for each task, the same source videos are used to generate both DVS and Retina events, while the downstream recurrent FireNet backbone, training schedule, and event-grid data augmentation procedure are kept fixed. Across tasks, only the task-specific target, output activation/head, and loss function are changed.

\subsection{Synthetic Data Generation and Task Selection}
Event-based vision has been successfully applied to Visual-Inertial Odometry (VIO) \cite{rebecq_high_2019, gallego_event-based_2022}, optical flow estimation \cite{bardow_simultaneous_2016, luo_efficient_2024} and object tracking \cite{shah_codedevents_2024}. To evaluate the benefits of the event streams, we utilise a neural network-based benchmark.

We use the synthetic data generation pipeline described in section \ref{ref:neural_upsampling} to generate event streams for both models. Although recent works have extended event simulation to include sensor noise modelling \cite{hu_v2e_nodate}, we deliberately exclude noise from this evaluation. Modelling the noise arising from transistor mismatch and thermal effects of SCAMP-5 is very non-trivial, and including noise for DVS but not for SCAMP-5 would introduce an unfair asymmetry. By comparing idealized noiseless DVS models with idealized noiseless Retina models, we aim to isolate the algorithmic/computational contribution of the Retina model relative to the DVS.

\begin{figure}[htbp]
  \centering
  \includegraphics[width=1.0\textwidth]{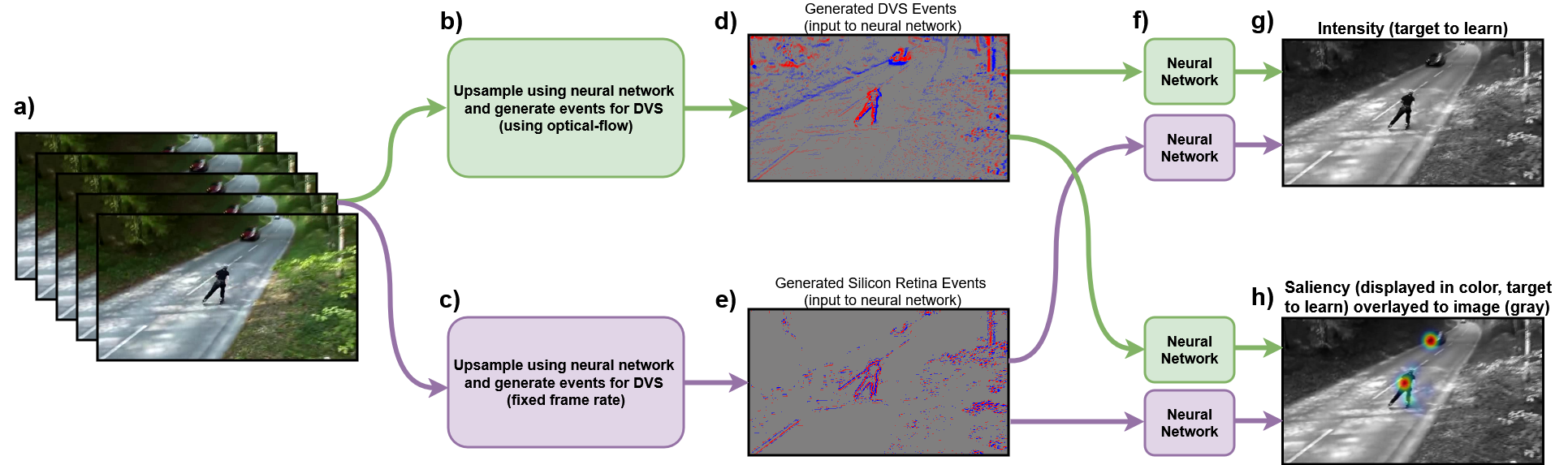}
  \caption{Overview of the experimental pipeline and evaluation benchmarks. (a) Input video sequence. (b, c) Frame upsampling stage utilizing neural networks: the DVS pipeline (b) uses an optical-flow-based approach~\cite{gehrig_video_2020}, whereas the Silicon Retina pipeline (c) uses a network to upsample frames to a fixed 200~FPS. (d, e) Event generation: the upsampled frames drive the simulators to produce spatiotemporal event streams $(x,y,p,t)$ for the standard DVS model (d) and our proposed Silicon Retina (e). (f) Evaluation methodology: we investigate how different event representations affect downstream neural network performance, keeping the network architecture and training procedure identical across comparisons. (g, h) Ground truth targets for the two evaluation tasks: (g) \textit{Video Intensity Reconstruction}, which aims to recover grayscale images from the event streams, and (h) \textit{Video Saliency Prediction}, where the model predicts human gaze attention maps (color overlay)~\cite{wang_revisiting_2021}.}
  \label{fig:neural_network_tasks}
\end{figure}

We evaluate performance on two distinct tasks (Fig. \ref{fig:neural_network_tasks}):
\begin{enumerate}
    \item \textbf{Video Intensity Reconstruction:} This task involves recovering original grayscale frames solely from the event stream. Since DVS events are generated via log-intensity differencing, reconstruction is a well-studied problem for DVS data. We investigate whether the complex retinal model preserves sufficient information to allow for similar reconstruction, effectively testing the compression of the encoding. 
    \item \textbf{Video Saliency Prediction:} This task involves predicting human attention (saliency) maps from the event stream. Unlike reconstruction, this requires extracting multiple cues for higher-level understanding rather than low-level pixel-intensity information. We use this task to evaluate the Retina model's ability to perform "filtering" and redundant-information rejection. 
\end{enumerate}


\subsection{Intensity reconstruction}
\label{sec:reconstruction}

\subsubsection{Task Definition}
A fundamental task for event-based vision is the reconstruction of full-intensity frames from the sparse event stream. For standard DVS sensors, this is a tractable problem: since events represent the changes in logarithmic intensity, this is essentially an integration task. This can be achieved via variational optimization \cite{bardow_simultaneous_2016, reinbacher_real-time_2016}, simple filters \cite{scheerlinck_continuous-time_2018} or lightweight neural networks \cite{rebecq_events--video_2019, scheerlinck_fast_2020}.

We selected this task because we believe it serves as a strong measure of encoding capacity. If the sensor output captures the complete visual signal, perfect reconstruction of the original signal should be possible. We hypothesize that the silicon retina model will struggle with this task compared to DVS. The OPL stage functions as a spatio-temporal band-pass filter, removing the intensity component to prioritise edges and movements. In this experiment, we want to quantify this information loss. 

\subsubsection{Experimental Setup}
We utilize the FireNet architecture proposed by Scheerlinck et al. \cite{scheerlinck_fast_2020} due to its high efficiency, low parameter count ($\approx$ 38k) and demonstrated suitability for event-based intensity reconstruction. We train the network on the Event-MS-COCO dataset \cite{scheerlinck_fast_2020}, which consists of static images (from MS-COCO) mapped to a 3D plane and subjected to a random 6-DOF camera motion.

To simulate the sensor data, we modify the pipeline from \cite{gehrig_video_2020}: 
\begin{itemize}
    \item \textbf{DVS Simulation:} We use the approach proposed by Gehrig et al. \cite{gehrig_video_2020} of upsampling images to achieve sub-pixel optical flow, generating events based on log-intensity changes.
    \item \textbf{Retina Simulation:} Unlike DVS, the retina model relies heavily on temporal dynamics (e.g. filter decay rates, solving differential equations). Therefore, we upsample the dataset to a fixed temporal resolution (200 FPS) before processing it with the Retina CUDA kernels.
\end{itemize}

\subsubsection{Model Optimization}
To ensure a fair comparison, we optimized hyperparameters of the Silicon Retina Model. The digital simulation framework of the biological model contains over 30 tunable parameters, such as filter formulations, filter coefficients, and kernel sizes. We selected few key parameters governing LIF dynamics and low-pass filter time constants and manually tweaked the pipeline. 

We trained FireNet for 150--200 epochs using the Adam optimiser and a perceptual similarity loss function (LPIPS) \cite{zhang_unreasonable_2018} and temporal consistency loss using predicted optical flow \cite{gehrig_video_2020}. The learning rate was scheduled with a decay factor ($\gamma=0.9$) every 10 epochs.

\subsubsection{Results and Discussion}

\begin{figure}[htbp]
  \centering  \includegraphics[width=1.0\textwidth]{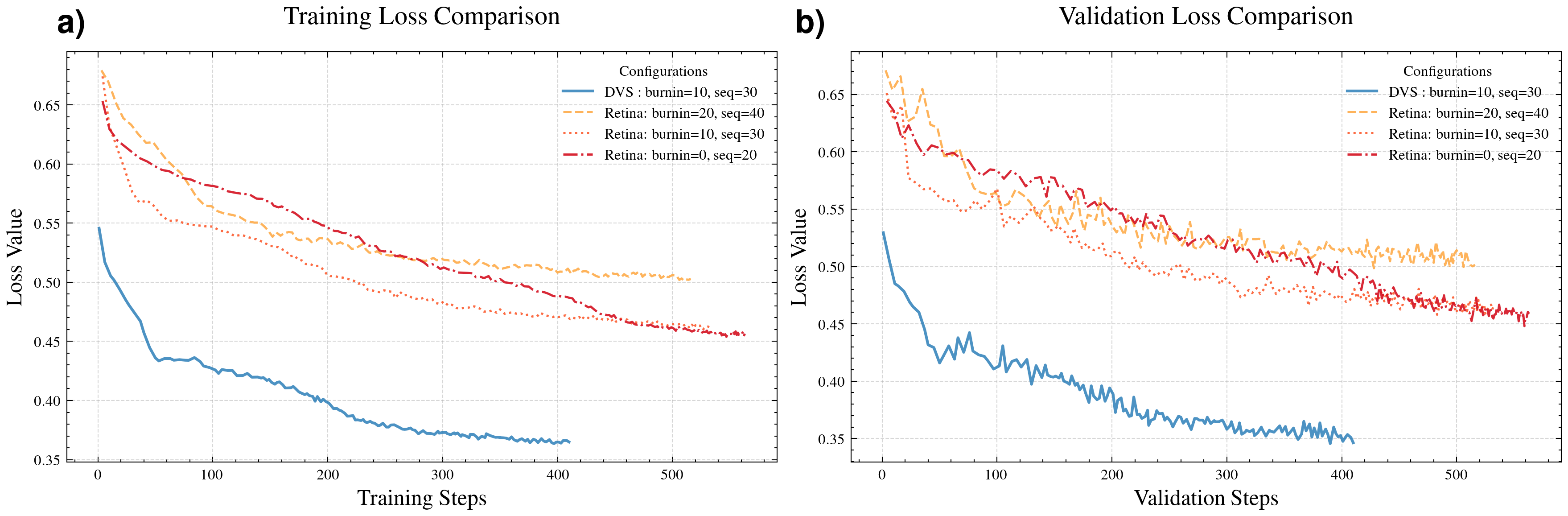}
  \caption{Performance on Video Intensity Reconstruction. (a) Training loss and (b) Validation loss over epochs. The standard DVS model consistently achieves lower reconstruction error, confirming that the Retina model discards absolute intensity information to prioritize edge and motion features. The "seq" value defined the total length of video in frames, while the "burnin" period represents the number of first $N$ frames, for which we don't compute the loss.}
  \label{fig:intensity}
\end{figure}

The training and validation loss curves are presented in Fig. \ref{fig:intensity}. Unlike DVS, the Retina model has significant internal state (filter histories, membrane states) that must settle  before any meaningful output is produced. To account for this, we implemented a "burn-in" stage during training: sampled video sequences are run through the model for a fixed number of initial frames to initialize the retinal states without computing the loss. After the burn-in stage, the network is tasked to reconstruct next fixed number of intensity frames. Fig. \ref{fig:intensity} compares DVS performance against several Retina configurations with varying burn-in durations. 

As hypothesized, the network trained on standard DVS events significantly outperforms the Retina-based models. The DVS baseline achieves a best loss value of 0.345 whereas the best retina configuration achieves $\approx $ 0.448 (configurations with burnin set to 0 and 10). This represents a performance gap of $\approx 0.103$, or a 30\% relative increase in reconstruction error for the Retina model. 

This performance gap confirms the theoretical properties of the retinal model. The OPL layer functions as a spatial high-pass filter, effectively discarding the low-frequency absolute intensity information required for accurate image reconstruction. While this makes intensity recovery an ill-posed problem for the Silicon Retina, it is not a system failure. Rather, it demonstrates that the sensor successfully performs desired information filtering — removing static intensity data to prioritize the edge and motion features relevant for dynamic vision.

\subsection{Video Saliency Prediction}
\label{sec:saliency}

\subsubsection{Task Definition and Motivation}
While the previous experiment demonstrated that the Silicon Retina struggles to preserve absolute intensity, we argue that for many biological and robotic tasks, intensity reconstruction is computationally inefficient and unnecessary. It involves sparse events and interpolating it back to dense frames for easy human interpretation. While this is beneficial in, for example, high-speed video reconstruction \cite{rebecq_high_2019}, this "information interpolation" removes the benefits of sparsity processing. 

In contrast, Video Saliency Prediction represents "information distillation". The goal is to identify and focus on the most relevant regions of the scene, discarding the noise or irrelevant information. This aligns closely with the functional role of the biological retina, which processes visual signal before transmission to the brain. Saliency prediction is inherently more complex than reconstruction, as ground truth is not defined by simple pixel mechanics but by human gaze behavior. This often includes high-level semantic understanding: for example, in one of the dataset videos, the saliency is focused on the instructor doing various movements. 

\subsubsection{Dataset Curation and Processing}
To evaluate this task, we utilize the Dynamic Human Fixation 1K (DHF1K) dataset \cite{wang_revisiting_2021}, which contains diverse video sequences paired with human eye-fixation maps. However adapting this dataset for high-speed event simulation presented specific challenges. The provided videos exhibit extreme variance in duration (ranging from 900 to 5000 frames after upsampling). These long sequences present significant challenge particularly given the memory constraints of batch training.

Therefore, we selected a subset of the data by filtering out sequences with excessive duration. Furthermore, we also removed the videos which had unreliable optical flow : for example in a single data point where multiple, unrelated videos are showed. This resulted in a final dataset of 541 videos, split into 441 for training and 100 validation, stratified by their duration. The sequences were processed using the same synthetic pipeline described in Section \ref{sec:reconstruction}.

\subsubsection{Experimental Setup}
We trained a lightweight Convolutional Neural Network (a FireNet with larger number of convolutional filters then original,  $\approx$100k parameters
) to predict saliency maps from the event streams. The loss function was a weighted combination of Kullback-Leibler (KL) divergence and the Pearson Correlation Coefficient (CC) (similarly to the dataset \cite{wang_revisiting_2021} ).

Our goal was not to surpass state-of-the-art saliency models, but to perform a relative comparison between the retina and DVS events using a fixed, proven and efficient neural network architecture. 

\subsubsection{Results and Discussion}
\begin{figure}[htbp]
  \centering
  \includegraphics[width=1.0\textwidth]{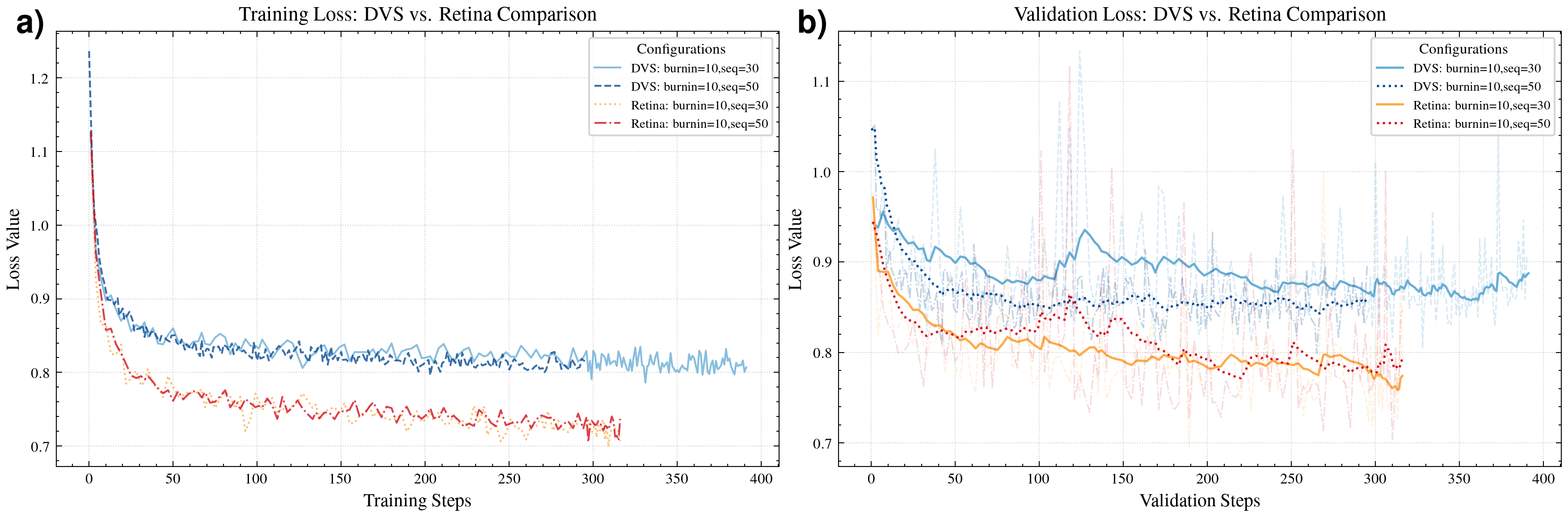}
  \caption{Performance on Video Saliency Prediction. (a) Training loss and (b) Validation loss over epochs ( the measurements were quite noisy, most likely to very small batch size of 2). Unlike intensity reconstruction, the Retina model achieves lower loss than the DVS baseline, despite generating significantly fewer events.}
  \label{fig:saliency_overall}
\end{figure}

The plot for training and validation results are in Fig. \ref{fig:saliency_overall}. In contrast to the intensity reconstruction, the Silicon Retina outperforms the standard DVS. In sparsity experiments, we define the event density using the average voxel-grid occupancy. For an event tensor $G \in \mathbb{R}^{T \times B \times H \times W}$, we define it as the fraction of time-bin/pixel entries whose absolute value is non-zero, averaged over all sequences in the dataset.

\begin{itemize}
    \item Performance: The Retina model achieves a best validation loss of 0.695, compared to 0.799 for the DVS baseline. This represents a robust improvement in predictive capability.
    \item Sparsity Efficiency: the performance is achieved with significantly less data. The Retina model generates events with an average grid occupancy of $\approx 5.7\%$, whereas the DVS baseline has $\approx 10.8\% $. The parameters of retina model are very similar to the ones used in intensity reconstruction, while the DVS is the same. Increasing sparsity of retina and decreasing sparsity of DVS is discussed in section \ref{sec:ablation}, however, for reasonable values, the Retina overall outperforms the DVS.    
\end{itemize}

These results support the hypothesis that the bio-inspired model suppresses redundant or weakly informative events, producing a filtered event stream that is more useful for saliency prediction. Retina events remove information needed for photometric reconstruction, as shown by the higher intensity-reconstruction loss. However, for saliency prediction, this representation preserves or enhances task specific information, allowing the network to achieve lower validation loss while emitting substantially fewer events than the DVS baseline.

\subsection{Effect of Retinal Model Parameters on Saliency Prediction}
\label{sec:ablation}

The results in Section~\ref{sec:saliency} show that Retina events can improve saliency prediction while producing a sparser event stream than the DVS baseline. However, unlike a standard DVS, the silicon retina model is not controlled by a single threshold parameter. The SCAMP-5 implementation exposes approximately 20 tunable parameters, while the digital simulator supports more than 30, including temporal filter coefficients, feedback gains, spatial kernel sizes, and LIF neuron constants. This makes it important to understand how different parameters and modelling choices affect the event representation, the resulting event density, and the downstream saliency prediction performance.

In this section, we therefore study a small set of parameters and design choices that are most relevant to the behaviour of the retina model and its implementation on a PPA. Specifically, we examine the contrast gain control formulation, the relationship between event density and saliency performance, the use of simplified high-pass filtering and the effect of the kernel-size in the contrast gain control. These experiments allow us to identify which aspects of the model are responsible for the observed sparsity, what performance trade-off and hardware-oriented approximations can be made without removing the advantage of Retina events over standard DVS events.

\subsubsection{Feedback formulation}
One of the key features of the model proposed by Wohrer \cite{wohrer_virtual_2009} is the inclusion of non-linear contrast adaptation mechanism which in its initial formulation, has quadratic form, as in equation \ref{qfunc}. 

\begin{figure}
  \centering  \includegraphics[width=0.6\textwidth]{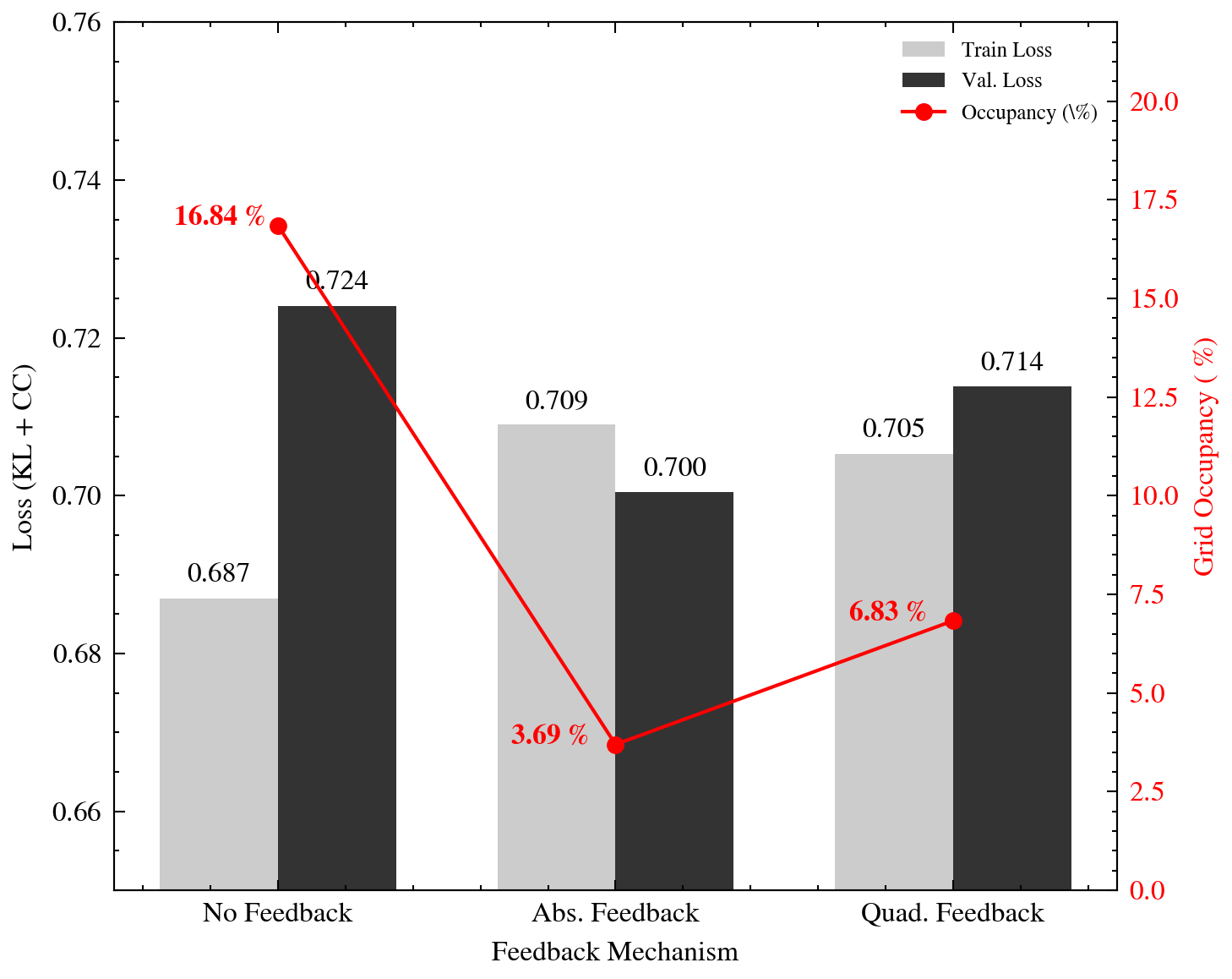}
  \caption{Performance (Training and Validation loss) and grid occupancy of FIRENet on video saliency reconstruction dependent on different feedback modes.}
  \label{fig:feedback_importance}
\end{figure}

In our SCAMP-5 implementation we simplified this equation, due to hardware constraints. The equation uses absolute value rather than quadratic formulation, as in equation \ref{eq:feedback_relaxed}. In this subsection we study how this mechanism affects the performance of the neural network. We consider 3 cases : 
\begin{enumerate}
    \item \textbf{No Feedback} : turning this stage off (feeding the outer plexiform layer current directly the retinal ganglion cell),
    \item \textbf{Abs. Feedback} : absolute value formulation ( equation \ref{eq:feedback_relaxed}),
    \item \textbf{Quad. Feedback} : quadratic formulation (equation \ref{qfunc})
\end{enumerate}
Most of the parameters are the same, with the main difference being $g_{A}^0$ and $\lambda_{A}$ vary in quadratic and absolute value cases. The results can be seen in Fig. \ref{fig:feedback_importance}.

An important observation here, is that the feedback limits the amount of events significantly: the grid occupancy with no feedback is $\approx 16.8 \%$, whereas the absolute value feedback reduces grid occupancy to less than 4\%, and the quadratic feedback yields 6.83 \%. When evaluating on the validation portion of data, the absolute-value feedback formulation slightly outperforms the other variants (0.700 against 0.714 for Quadratic feedback and 0.724 for No Feedback).

\subsubsection{High pass formulation}

A major challenge in implementing recursive temporal filters on analog processing arrays like SCAMP-5 is the accumulation of noise. Standard Infinite Impulse Response (IIR) formulations require multiple sequential analog operations per pixel. Specifically, the analog division operation is highly susceptible to noise amplification. When cascaded through the Outer Plexiform Layer (OPL) and Contrast Gain Control stages, this noise accumulation severely degrades the visual signal quality, rendering complex IIR implementations impractical on this hardware.

Consequently, our proposed pipeline utilizes a hardware-efficient "Simple Differencing" scheme, as it preserves most of the signal. 
\begin{equation}
A_{HPF1}[t] = B[t] - B[t-1]
\label{eq:hpf1}
\end{equation}

We contrast this with a standard tunable IIR high-pass formulation, presented in equation \ref{eq:standard_high_pass}. 

To validate our design choice and determine if utilizing using simple frame difference scheme (referred to as "Simple") incurs a significant performance penalty compared to standard high pass formulation ( referred to as "IIR"), we evaluated four configurations across the two temporal filtering stages (OPL, Ganglion cell): (Simple, Simple), (Simple, IIR), (IIR, Simple), and (IIR, IIR). We assessed these configurations under two operating regimes: a high occupancy regime ($p_{2} = 1.2$) and a low occupancy regime ($p_{2} = 1.0$).

\begin{figure}[htbp]
\centering
\includegraphics[width=1.0\textwidth]{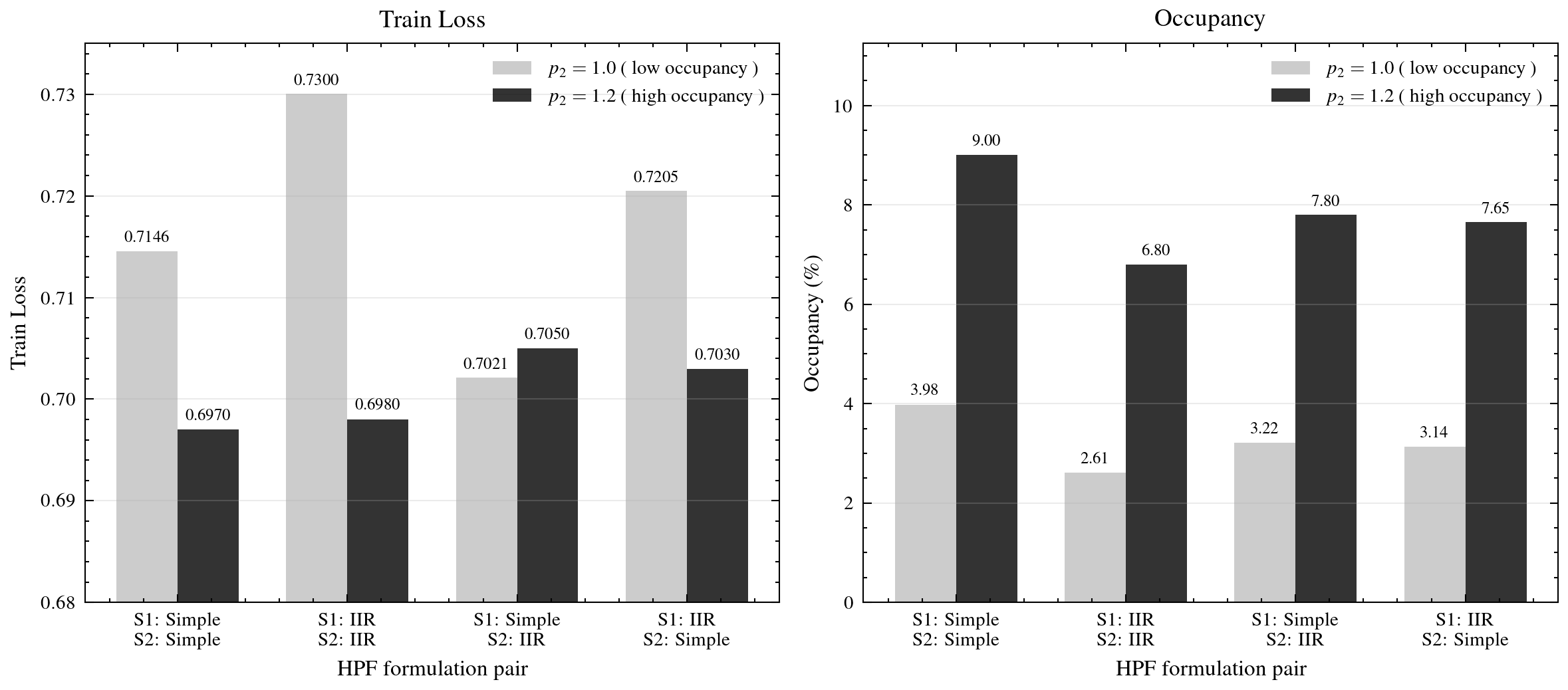}
\caption{Impact of high-pass filter formulations on training loss and grid occupancy. Performance is evaluated across two grid occupancy regimes (low $p_2$ and high $p_2$) mapping the filter choice across the OPL (Stage 1) and Ganglion Cell (Stage 2).}
\label{fig:HPF_formulations}
\end{figure}

The results are summarized in Fig.~\ref{fig:HPF_formulations}. The key findings are : 
\begin{itemize}
\item \textbf{High Occupancy Regime:} When the event density is sufficient,temporal filter formulation does not affect network performance . The maximum loss difference between configurations is negligible ($\approx 0.008$), indicating that the precise mathematical shape of the temporal transient is irrelevant when the overall event volume is high.
\item \textbf{Low Occupancy Regime:} At lower data rates, the network is sensitive to the filter formulation, primarily because the filter choice directly alters the baseline event generation rate. While the mixed configuration (Simple, IIR) achieves the absolute lowest loss (0.702), the fully simplified hardware baseline (Simple, Simple) remains highly competitive (0.714) and explicitly outperforms the full IIR configuration (IIR, IIR), which scores 0.720.
\end{itemize}

These results confirm that the simplified frame differencing formulation does not impose a critical bottleneck on downstream saliency prediction. It provides sufficient temporal acuity while circumventing the noise accumulation issues inherent to complex recursive filtering on an analog array.

\subsubsection{Kernel size}

\begin{figure}
   \centering
\includegraphics[width=0.6\textwidth]{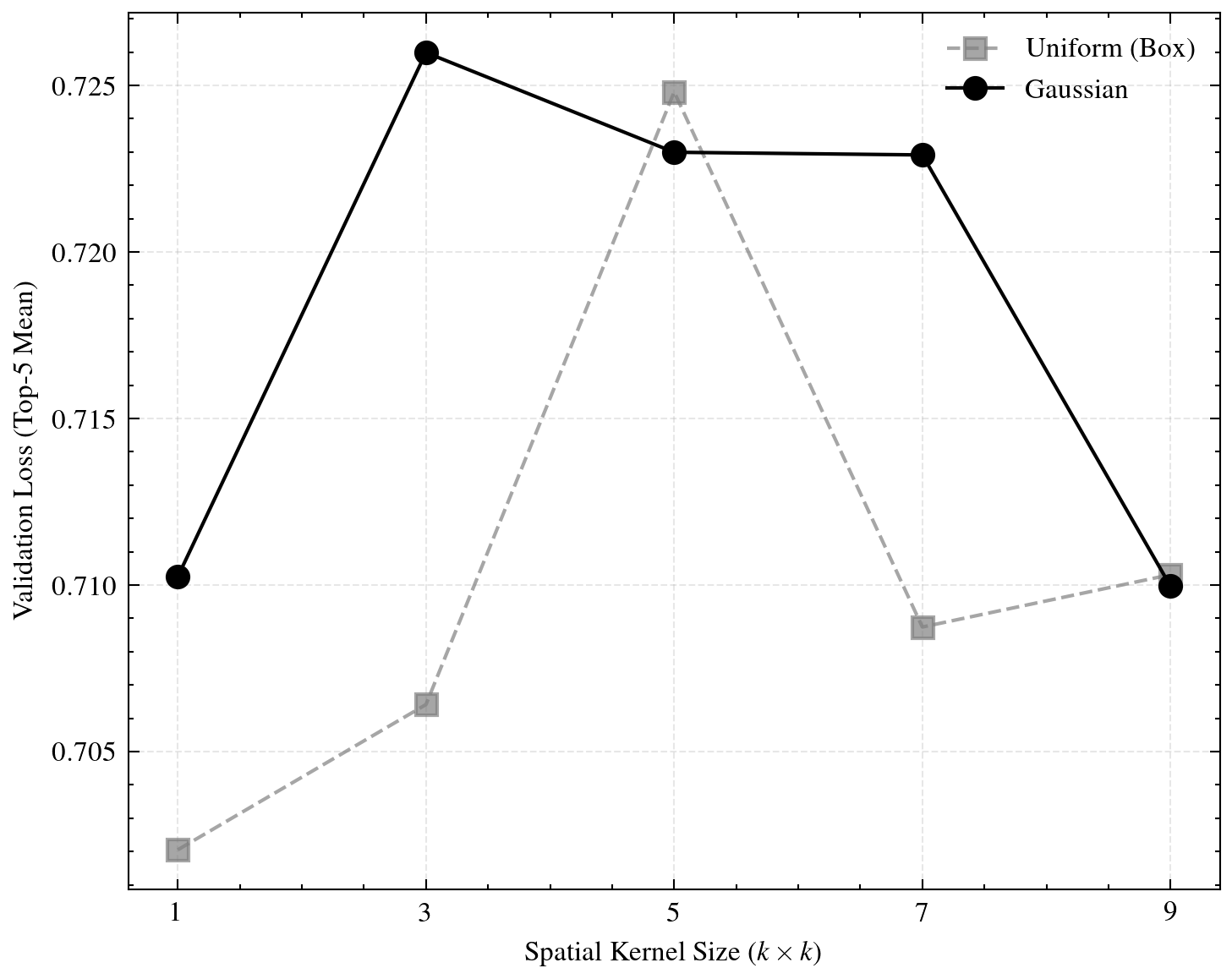}
\caption{Impact of spatial kernel size on neural network performance during the contrast gain control stage. Validation loss remains stable (0.70--0.72) across configurations, demonstrating that lightweight, localized spatial filtering is sufficient and complex large-kernel operations are unnecessary for hardware implementation.}
\label{fig:kernel_size}
\end{figure}

One of the key differences between the silicon retina and DVS is the spatial aspect of the retina. In this subsection we study the performance of the neural network against the size of the kernel in contrast gain control stage. The results are summarized in Fig. \ref{fig:kernel_size}
. Contrary to the expectation that wider biological receptive fields would significantly enhance performance, the performance of the system depends only weakly on the kernel size. The validation loss remains stable within a narrow band (0.70 -- 0.72) across all configurations.
 
These findings are highly favorable for hardware implementation. They suggest that the lightweight spatial filtering (or even purely local pixel processing) is sufficient to achieve better performance, without the need for complex, large-kernel operations.

\newpage
\subsubsection{Retina and DVS sparsity}

Silicon retina has the ability to carefully tune the event rate. This can be achieved by tuning the parameters of LIF neurons and could be tuned adaptively. An easy and direct method in our model is modulating the coefficient $p_2$ in the Ganglion Layer (Equation \ref{eq:LIF_int}), which scales the current injected into the Leaky Integrate and Fire (LIF) neurons

\begin{figure}[htbp]
  \centering
  \includegraphics[width=0.75\textwidth]{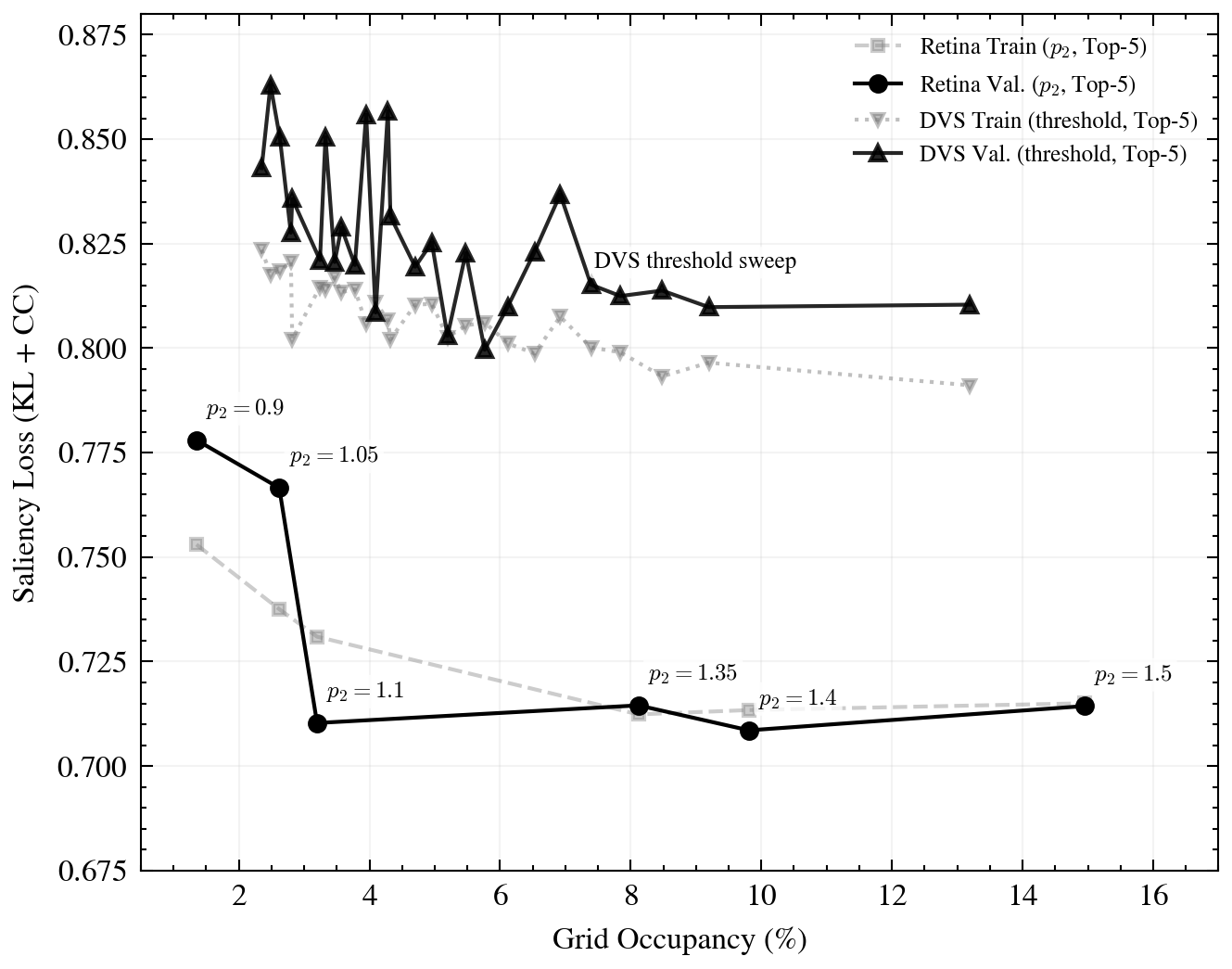}
  \caption{Saliency prediction performance as a function of grid occupancy. The plot illustrates the Sparsity-Performance tradeoff, where the data rate is modulated by the coupling parameter $p_2$. The secondary top axis indicates the specific $p_2$ coefficient corresponding to the measured grid occupancy.}
  \label{fig:performance_vs_sparsity}
\end{figure}

In this experiment, we investigate the relationship between the grid occupancy and saliency prediction performance by sweeping $p_2$. The results are presented in Fig. \ref{fig:performance_vs_sparsity}. We evaluate a range of sparsity levels from $\approx 1\%$ to $\approx 15\%$.

As illustrated in Fig. \ref{fig:performance_vs_sparsity}, the system exhibits following behaviour: 
\begin{enumerate}
    \item The optimal performance is obtained at $\approx 3.5 - 10 \%$. We observe a sharp performance improvement for grid occupancies higher than $\approx 3.5 \%$. The validation loss stabilizes in this region (when keeping other parameters fixed).
    \item The training loss continues to decrease monotonically as data density increased, validation performance plateaus or slightly degrades, suggesting that additional events contribute primarily redundant information or noise.
\end{enumerate}

The Retina model consistently outperforms the obtained DVS baseline across all density levels. For instance, at $\approx 10\%$ occupancy—which corresponds to the default threshold configuration for DVS in~\cite{gehrig_video_2020}—the Retina model achieves a loss of $\approx 0.70$, whereas the DVS baseline sits at $\approx 0.80$. 

\subsection{DVS event generation on silicon retina upsampling}

In the previous experiments, the Silicon Retina model demonstrated lower loss than the standard DVS baseline. However, the data generation pipelines for the two models differed: the standard DVS benchmark utilized an optical-flow-based upsampling approach \cite{gehrig_video_2020} to simulate asynchronous operation of the DVS, whereas the Silicon Retina relied on a fixed 200 FPS upsampling to mimic the time-stepped execution of the SCAMP-5 hardware.

This discrepancy raises an important question, especially in the context of obtained results: does the Silicon Retina's performance gain come from its bio-inspired architecture, or does the fixed-rate upsampling create an "easier" data distribution for the saliency network? 

We perform the following experiment to answer that question: we fed the exact same fixed-framerate (200 FPS) input data used by the Silicon Retina into the DVS simulator. This effectively isolates the sensor model (DVS vs. Silicon Retina) as the sole difference between the generated event streams.

\begin{figure}[htbp]
    \centering
    \includegraphics[width=1.0\linewidth]{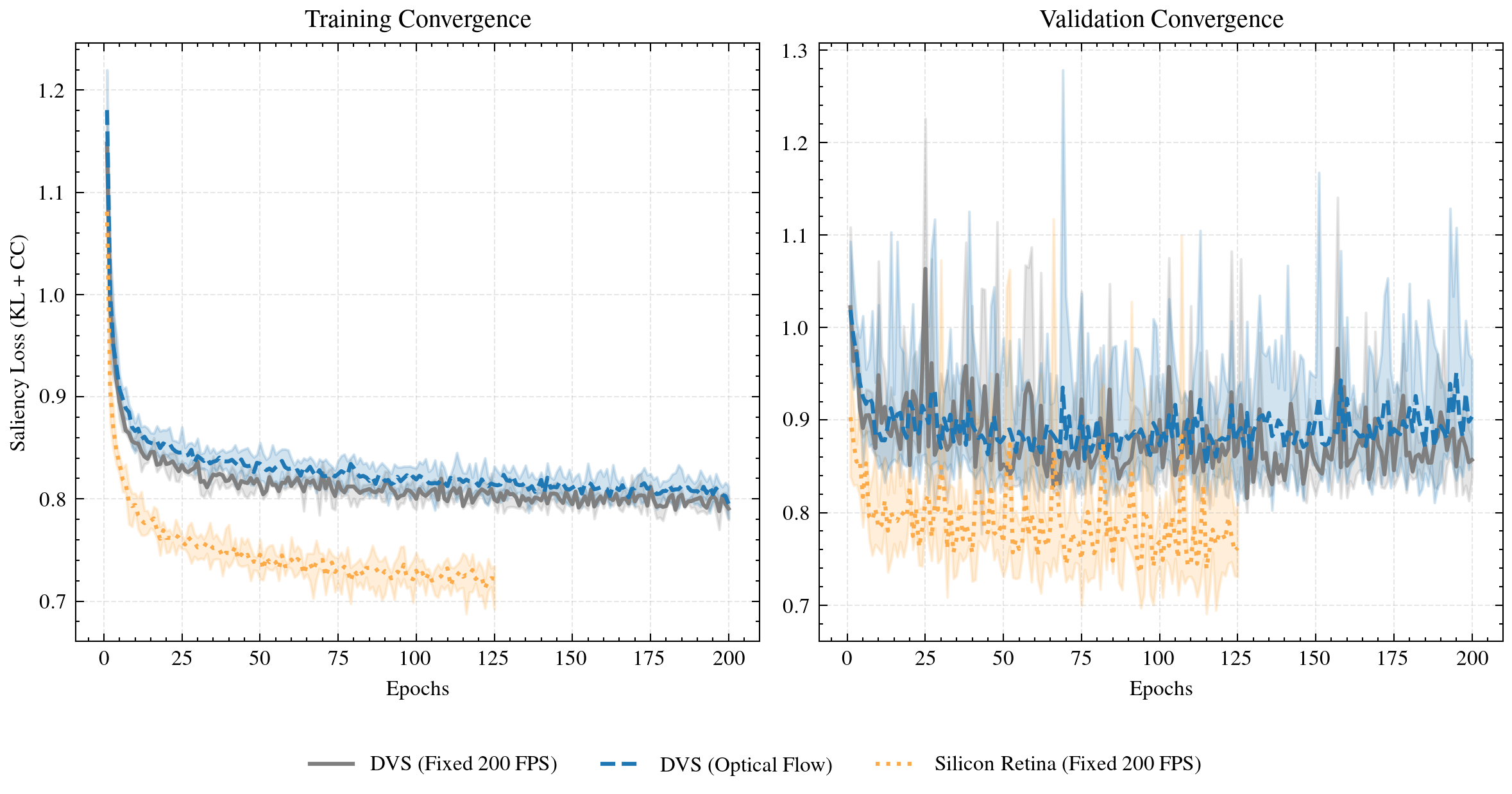}
    \caption{Training and validation convergence curves isolating the upsampling methodology. Forcing the DVS model to use the fixed 200~FPS data (gray trace) do not affect the performance signficiantly. The Silicon Retina (orange trace) achieves the lowest loss, validating the performance of the retinal model.}
    \label{fig:upsampling_comparison}
\end{figure}

The convergence results are presented in Fig. \ref{fig:upsampling_comparison}. Curves show the mean and shaded min-max range across many trainings: "DVS (Optical Flow)" over 7 different seeds, "DVS (Fixed 200 FPS)" over 3 different seeds and "Silicon Retina (Fixed 200 FPS)" over 3 different seeds. The DVS model evaluated on the fixed-framerate data achieves a very similar loss than the DVS baseline utilizing optical flow. While the fixed-framerate DVS model achieve slightly lower loss ($\approx 1\%$), the retina model achieves significantly ($\approx 12\%$ lower loss). This confirms that the fixed 200 FPS upsampling used by Silicon Retina model does not provide an unfair advantage for the saliency task. Silicon Retina (orange) achieves the lowest loss across all evaluated configurations, indicating that advantage comes from the retinal processing model.

By demonstrating that the fixed-rate upsampling do not affects the performance significantly, this experiment confirms that the Silicon Retina's better performance is due to its internal processing mechanisms. 

\section{Discussion and Conclusions}
\label{sec:conclusion}

In this work, we investigated a bio-inspired alternative to standard event-based sensing, implementing a multistage retina model on Pixel Processor Array (PPA). By leveraging the parallel processing capabilities and near-photodiode computing of the SCAMP-5 system, we incorporated important biological mechanisms  - specifically center-surround spatial filtering and contrast gain control - directly at the focal plane. To facilitate further research, we also introduced a GPU-based simulation framework capable of generating these retinal events from video data or high-speed camera inputs. 

We evaluated the implications of this added complexity on two fundamental tasks: Video Intensity Reconstruction and Video Saliency Prediction. 
In the reconstruction task, the Silicon Retina model struggled to preserve luminance information compared to the standard DVS baseline. This confirms that the Outer Plexiform Layer effectively filters absolute intensity data, and the network was not capable of reconstruction. However this loss of photometric fidelity proved somewhat advantageous for high-level semantic processing. In the Video Saliency Prediction task, the Silicon Retina model demonstrated improved performance over the DVS baseline, achieving a 13\% relative reduction in validation loss.

In our experimentation, the Silicon Retina achieved this lower loss while generating approximately 47\% fewer events than the standard DVS baseline. Our retina parameter studies confirmed that the bio-inspired Gain Control mechanism is the main mechanism of the data reduction. Furthermore, we validated that hardware-efficient approximations of biological functions, such as absolute-value feedback and simple differencing filters yield good performance comparable to their more complex mathematical counterparts. 


While standard event cameras are highly effective for a wide range of applications, our results suggest that for some applications, there are benefits of models which go beyond simple temporal difference. By mimicking the retina's ability to pre-filter and distill visual information, we can achieve a favorable trade-off between data sparsity and predictive performance.

\section{Acknowledgment}
This work was initiated during the Bangalore Neuromorphic Engineering Workshop 2025. The authors would like to acknowledge the assistance given by Research IT and the use of the Computational Shared Facility at The University of Manchester for providing high-performance computing resources used for the simulations reported in this work.

\section*{Funding}
Maciej Lewandowski was supported by the University of Manchester through a Dean's Doctoral Award. This work was supported by the University of Manchester – IISc Research Fund (UoM/IISc Research Fund).

\section*{Data availability}
This study used the existing Event-MS-COCO and DHF1K datasets, described in Refs.~\cite{scheerlinck_fast_2020,wang_revisiting_2021}.
Instructions for obtaining and preprocessing these datasets are provided in the project repository. Numerical results underlying the reported plots, together with the corresponding analysis notebooks, are available at \url{https://github.com/mlewandowski0/silicon-retina}. Synthetic event streams are generated from the source datasets using the released simulation code and configurations.

\section*{Code availability}
The SCAMP-5 implementation, silicon retina simulation code, training and evaluation scripts, experiment configurations, and figure-generation notebooks are available at \url{https://github.com/mlewandowski0/silicon-retina}.

\section*{Author contributions}
Maciej Lewandowski contributed to the conceptualization, developed the methodology, implemented the hardware and software, conducted all experiments, analysed the results, prepared the figures, and wrote the original manuscript. Prince Philip contributed to the conceptualization and manuscript development. Alexandre Marcireau contributed to the experimental design and manuscript development. Andr\'e van Schaik and Piotr Dudek supervised the research, contributed to manuscript development. Chetan Singh Thakur contributed to manuscript development. All authors reviewed and revised the manuscript.

\printbibliography

\end{document}